\documentclass[symmetry,article,accept,moreauthors,pdftex]{Definitions/mdpi} 

\firstpage{1} 
\pubvolume{xx}
\issuenum{1}
\articlenumber{5}
\pubyear{2019}
\copyrightyear{2019}
\history{Received: date; Accepted: date; Published: date}

\usepackage{amssymb}
\usepackage{amsfonts}
\usepackage{ragged2e}
\usepackage[official]{eurosym}

\usepackage{url}
\usepackage{amsmath,amsfonts,amssymb,amsthm}
\usepackage{bm,bbm,bbold,braket,xcolor,mathrsfs}
\usepackage{graphicx}
\usepackage{array}
\usepackage{float}
\usepackage{color}
\usepackage{footnote}
\usepackage{hyperref}
\hypersetup{linktocpage,colorlinks=true,allcolors=blue}

\Title{Cosmic Ray Extremely Distributed Observatory}

\newcommand\namefamily{\fontfamily{lmvtt}\selectfont}
\newcommand\narrowuscore{\nobreak\hskip0pt plus0.01667pt\vbox{\hrule width.3em height-.0860em depth.1640em}\penalty\binoppenalty\hskip0pt plus0.01667pt}
\newcommand\usenarrowuscore{\let\_\narrowuscore}
\newcommand{\textname}[1]{{\namefamily\usenarrowuscore#1}}
\newcommand{\qn}[2]{\textname{#1:}\penalty\binoppenalty\hskip0pt\textname{#2}}

\definecolor{sangria}{rgb}{0.57, 0.0, 0.04}

\definecolor{electricviolet}{rgb}{0.56, 0.0, 1.00}

\newcommand{\Lag}{\mathscr{L}}

\newcommand{\LBgcc}{\mathrm{g}/\mathrm{cm}^{3}}
\newcommand{\LBms}{\mathrm{ms}}
\newcommand{\LBerg}{\mathrm{erg}}
\newcommand{\LBmsun}{\mathrm{M}_{\odot}}
\newcommand{\LBmearth}{\mathrm{M}_{\oplus}}
\newcommand{\LBkm}{\mathrm{km}}
\newcommand{\LBergsec}{\mathrm{erg}/\mathrm{s}}
\newcommand{\LBsec}{\mathrm{s}}

\Author{Piotr Homola$^{1}$*\orcidA{}, Dmitriy Beznosko$^{2}$\orcidB{}, Gopal Bhatta$^{3}$, \L{}ukasz Bibrzycki$^{4}$\orcidJ{}, Michalina Borczy\'n{}ska$^{5}$, \L{}ukasz Bratek$^{6}$\orcidV{}, Nikolai Budnev$^{7}$, Dariusz Burakowski$^{8}$, David E. Alvarez-Castillo$^{1,9}$\orcidD{}, Kevin Almeida Cheminant$^{1}$, Aleksander \'Cwik\l{}a$^{10}$, Punsiri Dam-o$^{11}$, Niraj Dhital$^{12}$, Alan R. Duffy$^{13}$\orcidI{}, Piotr G\l{}ownia$^{6}$, Krzysztof Gorzkiewicz$^{1}$, Dariusz G\'ora$^{1}$, Alok C. Gupta$^{14}$, Zuzana Hl\'avkov\'a$^{15}$, Martin Homola$^{15}$\orcidC{}, Joanna Ja\l{}ocha$^{6}$, Robert Kami\'nski$^{1}$, Micha\l{} Karbowiak$^{16}$\orcidE{}, Marcin Kasztelan$^{17}$, Renata Kierepko$^{1}$, Marek Knap$^{8}$, P\'eter Kov\'acs$^{18}$, Szymon Kuli\'nski$^{5}$, Bartosz \L{}ozowski$^{19}$, Marek Magry\'s$^{20}$, Mikhail V. Medvedev$^{21,22}$, Justyna M\k{e}drala$^{23}$, Jerzy W. Mietelski$^{1}$, Justyna Miszczyk$^{1}$, Alona Mozgova$^{24}$, 
Antonio Napolitano$^{25}$\orcidZ{}, Vahab Nazari$^{1,9}$, Y. Jack Ng$^{26}$, 
Micha\l{} Nied\'zwiecki$^{10}$\orcidN{}, Cristina Oancea$^{27,28}$, Bogus\l{}aw Ogan$^{8}$, Gabriela Opi\l{}a$^{23}$, Krzysztof Oziomek$^{20}$, Maciej Pawlik$^{20,23}$, Marcin Piekarczyk$^{4}$\orcidX{}, Bo\.zena Poncyljusz$^{5}$, Jerzy Pryga$^{29}$, Mat\'ias Rosas$^{30}$, Krzysztof Rzecki$^{23}$\orcidR{}, Jilberto Zamora-Saa$^{31}$\orcidW{}, Katarzyna Smelcerz$^{10}$\orcidK{}, Karel Smolek$^{32}$, Weronika Stanek$^{23}$, Jaros\l{}aw Stasielak$^{1}$, S\l{}awomir Stuglik$^{1}$\orcidS{}, Jolanta Sulma$^{33}$, Oleksandr Sushchov$^{1}$, Manana Svanidze$^{34}$, Kyle Tam$^{35}$, Arman Tursunov$^{36}$\orcidT{}, Jos\'e M. Vaquero$^{37}$\orcidU{}, Tadeusz Wibig$^{16}\orcidY{}$, Krzysztof W. Wo\'zniak$^{1}$}

\AuthorNames{Piotr Homola, Dmitriy Beznosko, Gopal Bhatta, \L{}ukasz Bibrzycki, Michalina Borczy\'n{}ska, \L{}ukasz Bratek, Nikolai Budnev, Dariusz Burakowski, David E. Alvarez-Castillo, Kevin Almeida Cheminant, Aleksander \'Cwik\l{}a, Punsiri Dam-o, Niraj Dhital, Alan R. Duffy, Piotr G\l{}ownia, Krzysztof Gorzkiewicz, Dariusz G\'ora, Alok C. Gupta, Zuzana Hl\'avkov\'a, Martin Homola, Joanna Ja\l{}ocha, Robert Kami\'nski, Micha\l{} Karbowiak, Marcin Kasztelan, Renata Kierepko, Marek Knap, P\'eter Kov\'acs, Szymon Kuli\'nski, Bartosz \L{}ozowski, Marek Magry\'s, Mikhail V. Medvedev, Justyna M\k{e}drala, Jerzy W. Mietelski, Justyna Miszczyk, Alona Mozgova, Antonio Napolitano, Vahab Nazari, Y. Jack Ng, Micha\l{} Nied\'zwiecki, Cristina Oancea, Bogus\l{}aw Ogan, Gabriela Opi\l{}a, Krzysztof Oziomek, Maciej Pawlik, Marcin Piekarczyk, Bo\.zena Poncyliusz, Jerzy Pryga, Mat\'ias Rosas, Krzysztof Rzecki, Jilberto Zamora-Saa, Katarzyna Smelcerz, Karel Smolek, Weronika Stanek, Jaros\l{}aw Stasielak, S\l{}awomir Stuglik, Jolanta Sulma, Oleksandr Sushchov, Manana Svanidze, Kyle Tam, Arman Tursunov, Jos\'e M. Vaquero, Tadeusz Wibig, Krzysztof W. Wo\'zniak}

\address{%
$^{1}$ \quad Institute of Nuclear Physics Polish Academy of Sciences, Radzikowskiego 152, Krak\'ow, Poland\\
$^{2}$ \quad Clayton State University, Morrow, Georgia, USA\\
$^{3}$ \quad Astronomical Observatory, Jagiellonian University, Krak\'ow, Poland\\
$^{4}$ \quad Institute of Computer Science, Pedagogical University of Krakow, ul. Podchorążych 2, 30-084 Krak\'ow, Poland; lukasz.bibrzycki@up.krakow.pl (Ł.B.); marcin.piekarczyk@up.krakow.pl (M.P.)\\
$^{5}$ \quad University of Warsaw, Faculty of Physics, Poland\\
$^{6}$ \quad Institute of Physics, Faculty of Materials Engineering and Physics, Cracow University of Technology, Podchor\c{a}\.{z}ych 1, PL-30084 Krak\'{o}w, Poland\\
$^{7}$ \quad Irkutsk State University, Russia\\
$^{8}$ \quad Astroparticle physics amateur; buriaszany@gmail.com (D.B.); mpknap@wp.pl (M.Kn.); spinserwis@gmail.com (B.O.)\\
$^{9}$ \quad Bogoliubov Laboratory for Theoretical Physics,
Joint Institute for Nuclear Research, Joliot-Curie street 6, 141980 Dubna, Russia; alvarez@theor.jinr.ru\\
$^{10}$ \quad Department of Computer Science, Faculty of Computer Science and Telecommunications, Cracow University of Technology, ul. Warszawska 24, 31-155 Krak\'ow, Poland \\
$^{11}$ \quad School of Science, Walailak University, 222 Thasala, Nakhon si thammarat 80160, Thailand\\
$^{12}$ \quad St. Xavier's College, Maitighar, Kathmandu, Nepal\\
$^{13}$ \quad Centre for Astrophysics and Supercomputing, Swinburne University of Technology, Melbourne, Victoria 3122, Australia; aduffy@swin.edu.au \\
$^{14}$ \quad Aryabhatta Research Institue of Observational Sciences (ARIES), Manora Peak, Nainital 263001, India\\
$^{15}$ \quad Comenius University in Bratislava, Mlynsk\'{a} dolina, 842\,48 Bratislava, Slovakia\\
$^{16}$ \quad University of \L \'od\'z, Faculty of Physics and Applied Informatics, 90-236 
\L \'od\'z, Pomorska 149/153, Poland; t.wibig@gmail.com (T.W.)\\
$^{17}$ \quad National Centre for Nuclear Research, Andrzeja So\l{}tana 7, 05-400 Otwock-\'Swierk, Poland\\
$^{18}$ \quad Institute for Particle and Nuclear Physics, Wigner Research Centre for Physics, 1121 Budapest, Konkoly-Thege Mikl\'os \'ut 29-33, Hungary\\
$^{19}$ \quad University of Silesia in Katowice, Faculty of Natural Sciences, Poland\\
$^{20}$ \quad ACC Cyfronet AGH-UST, Krak\'ow, Poland\\
$^{21}$ \quad Department of Physics and Astronomy, University of Kansas, Lawrence, KS 66045, USA; medvedev@ku.edu\\
$^{22}$ \quad Laboratory for Nuclear Science, Massachusetts Institute of Technology, Cambridge, MA 02139, USA; medvedev@mit.edu\\
$^{23}$ \quad AGH University of Science and Technology, Krak\'ow, Poland\\
$^{24}$ \quad Astronomical Observatory of Taras Shevchenko National University of Kyiv, 04053 Kyiv, Ukraine\\
$^{25}$ \quad University of Napoli ``Parthenope'', Napoli, Italy\\
$^{26}$ \quad University of North Carolina at Chapel Hill, USA\\
$^{27}$ \quad ADVACAM, Prague, Czech Republic\\
$^{28}$ \quad University of Bucharest, Bucharest, Romania\\
$^{29}$ \quad Jagiellonian University, Krak\'ow, Poland\\
$^{30}$ \quad Institute of Secondary Education, Highschool No. 65, Montevideo, Uruguay\\
$^{31}$ \quad Departamento de Ciencias Fisicas, Universidad Andres Bello, Sazie 2212, Piso 7, Santiago, Chile; jilberto.zamora@unab.cl \\
$^{32}$ \quad Institute of Experimental and Applied Physics, Czech Technical University in Prague, Czech Republic\\
$^{33}$ \quad Publiczna Szko\l{}a Podstawowa im. \'Sw. Jadwigi Kr\'olowej w Rzezawie, Poland\\
$^{34}$ \quad E. Andronikashvili Institute of Physics under Tbilisi State University, Georgia\\
$^{35}$ \quad Waterloo Rocketry, University of Waterloo, Waterloo, Ontario, Canada\\
$^{36}$ \quad Research Centre for Theoretical Physics and Astrophysics, Institute of Physics, Silesian University in Opava, Bezru{\v c}ovo n{\'a}m. 13, CZ-74601 Opava, Czech Republic; arman.tursunov@physics.slu.cz\\
$^{37}$ \quad Department of Physics, University of Extremadura, M\'erida, Spain\\
}

\corres{Correspondence: Piotr.Homola@ifj.edu.pl; Tel.: +48-12-662-8341}

\abstract{The Cosmic Ray Extremely Distributed Observatory (CREDO) is a newly formed, global collaboration dedicated to observing and studying cosmic rays (CR) and cosmic ray ensembles (CRE): groups of a minimum of two CR with a common primary interaction vertex or the same parent particle. The CREDO program embraces testing known CR and CRE scenarios, and preparing to observe unexpected physics, it is also suitable for multi-messenger and multi-mission applications. Perfectly matched to CREDO capabilities, CRE could be formed both within classical models (e.g. as products of photon-photon interactions), and exotic scenarios (e.g. as results of decay of Super Heavy Dark Matter particles). Their fronts might be significantly extended in space and time, and they might include cosmic rays of energies spanning the whole cosmic ray energy spectrum, 
with a footprint composed of at least two extensive air showers with correlated arrival directions and arrival times. Since CRE are mostly expected to be spread over large areas and, because of the expected wide energy range of the contributing particles, CRE detection might only be feasible when using available cosmic ray infrastructure collectively, i.e. as a globally extended network of detectors. Thus, with this review article, the CREDO Collaboration invites the astroparticle physics community to actively join or to contribute to the research dedicated to CRE, and in particular to share any cosmic ray data useful for the specific CRE detection strategies.}


\begin{document}

\section*{Introduction}
\label{sec-intro}

While state-of-the-art cosmic ray research to date has been focused on the detection and analysis of cosmic particles observed through individual detectors or arrays, the correlated observations of cosmic rays on the global scale remain insufficiently explored, yet no less promising. This collaborative perspective could provide a deeper insight into the physical processes within energy ranges rarely considered, including the highest energies known.
Here we discuss a general approach to research dedicated to detecting and studying the astroparticle physics phenomena called Cosmic Ray Ensembles (CRE) defined as groups of a minimum of two correlated, be it spatially or temporally, cosmic rays with a common primary interaction vertex or the same parent particle. Such particles, constituents of CRE, are messengers of the primary physical processes – probes of the physics that happened even billions of years ago, at energies even millions of times higher than energies to which we can accelerate particles using man-made infrastructure. 


\begin{figure}[H]
\centering
\includegraphics[width=0.8\textwidth]{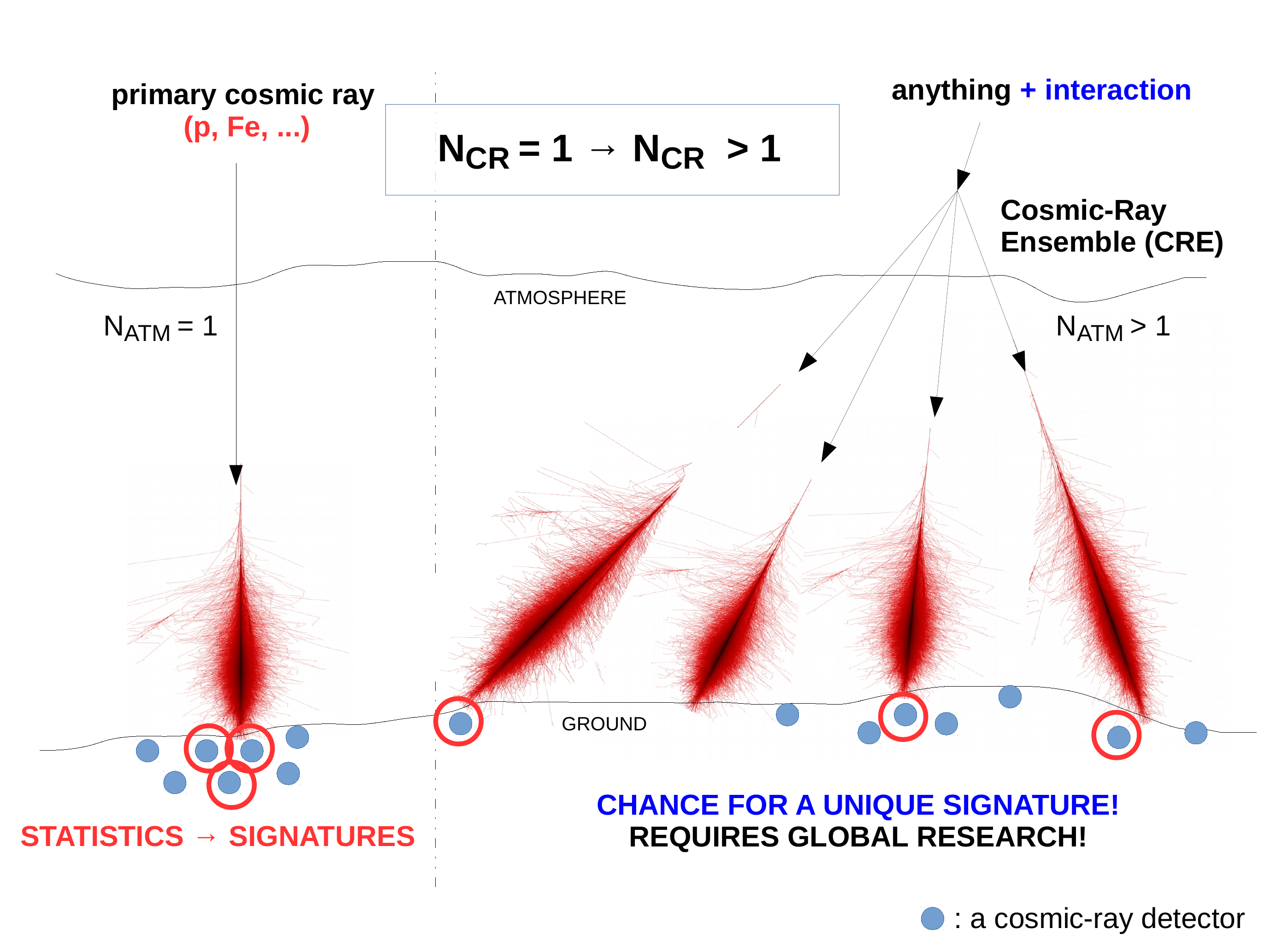}
\caption{Cosmic-Ray Ensembles: a novelty in cosmic-ray research and in multi-messenger astroparticle physics.}
\label{fig-novelty}
\end{figure}


\begin{justify}
Armed with the particle physics background telling us that cosmic ray particles are expected to interact with radiation and matter on their way through the Cosmos and give birth to CRE, we ask not whether CRE exist, but under which circumstances and with which conditions they can reach the Earth and be detected with the available or possible infrastructure. As illustrated in Fig. \ref{fig-novelty}, the signatures of CRE might be spread over very large surfaces which might make them hardly detectable by existing detector systems operating individually. On the other hand, if these detector systems operate under a planetary (and beyond) network, as proposed by The Cosmic-Ray Extremely Distributed Observatory (CREDO) \cite{Homola2018-gz,Gora2018-yx,Homola2019-xo}, the chances for detection of CRE will naturally increase. The components of CRE might have energies that practically span the whole cosmic-ray energy spectrum. Thus, all the cosmic ray detectors working in this range, beginning from smartphones (e.g. DECO \cite{Vandenbroucke2016-yj}, CRAYFIS \cite{Whiteson2016-vv}, CREDO Detector \cite{credo-detector,Bibrzycki2020-sy,Niedzwiecki2019arXiv190901929N}, Cosmic Ray App \cite{cosmic-ray-app}) and pocket size open-hardware scintillator detectors (e.g. Cosmic Watch \cite{Axani2018-pu,Schaub2018-mt} or CosmicPi \cite{cosmic-pi}), through numerous larger educational detectors and arrays (e.g. HiSPARC \cite{Colle2007-uf,Fokkema2012-so}, QuarkNet \cite{Bardeen2018-ua}, Showers of Knowledge \cite{Bychkov2012-dt}, CZELTA \cite{Smolek2008-ht}) to the professional infrastructure that receives or will receive cosmic rays as a signal or as a background Pierre Auger Observatory \cite{Auger_Collaboration2015-lz}, Telescope Array \cite{Kawai2008-uq}, JEM-EUSO \cite{Adams2015-ho,Adams2015-xk}, HAWC \cite{DeYoung2012-oz}, MAGIC \cite{Ferenc2005-ai}, H.E.S.S. \cite{Vasileiadis2005-wi}, VERITAS \cite{Krennrich2004-jq,Park2015-yj}, IceCube \cite{Aartsen2017-nm}, Baikal-GVD \cite{Avrorin2016-fp}, ANTARES Telescope \cite{Ageron2011-cr}, European Southern Observatory \cite{Franza1982-zd}, other astronomical observatories, underground observatories, accelerator experiments in the off-beam mode) could contribute to a common effort towards a hunt for CRE. Therefore it is not only desirable, but also feasible to put the CRE research into a routine implementation. So far, the experimental searches for cosmic-ray correlations have been realized on different scales with arrays of detectors located at schools and universities. However, all those efforts were dedicated to very specific scenarios, concerning mostly fragmentation of nuclei in background electromagnetic radiation fields \cite{Gerasimova1960-fh}, which limits the number of particles in the group (ensemble) to just a few. Some of these projects include e.g. CHICOS \cite{Carlson2005-li} in the U.S., ALTA \cite{Pospisil2009-cv} in Canada, GELATICA \citep{Svanidze2011-ac} in Georgia, EEE \cite{Abbrescia2016-fv} in Italy, LAAS \cite{Ochi2003-ww,Matsumoto2019-dd} in Japan, and the aforementioned CZELTA in Czech Republic. Time correlation of registered showers was studied at distances from 100 m to 7000 km, and in some cases evidence for unexpected coincidences have been found. However, these were without any convincing follow-up studies and data taking campaigns, which is hard without global coordination. Only very recently the idea of looking for large scale correlations in a general and global way took shape with the CREDO Collaboration, formalized in September 2019 \cite{credo_eurek}. CREDO is meant to be a multi-technique (different detector types) and doubly open (for both data upload and offering access) infrastructure enabling global research programs concerning radiation (both cosmic and terrestrial), with a number of multi-messenger, multi-mission and transdisciplinary opportunities. With this review article we invite the community to both benefit from the openness of CREDO and to contribute to its program with the research dedicated to a general search for ensembles of cosmic rays, especially photons of different energies. Since the concept of CREDO assumes the openness for independent data streams, it is expected that the projects mentioned above, as well as the other cosmic ray infrastructure including private, widely spread detectors such as smartphones, will have a direct interest in connecting to the global CREDO system, thus reinforcing its scientific attractiveness, and chances of individual research programs. 
\end{justify}\par

\begin{justify}
As this article is meant to serve as a review of the current status and perspectives of CREDO and the research related to CRE, its structure follows the scientific and logical roadmap to observing and studying cosmic ray ensembles. We begin with a general description of the field in Section \ref{sec-phil} ``Foundations of the CREDO methodology'' and \ref{sec-land} ``CREDO within the cosmic ray landscape'', and theoretical modeling of CRE sources in Section \ref{sec-uhecr} ``UHECR sources and cosmic ray ensembles''. Then we present and discuss example CRE scenarios with an emphasis on the simulations of propagation of the CRE components through the Cosmos and within the Earth's atmosphere (Section \ref{sec-sims} ``CRE simulations''), describe the status of the observational efforts (Sections \ref{sec-detectors} ``CREDO detectors: cloud of clouds'', \ref{sec-data} ``Data management and analysis'',
and \ref{sec-citizens} ``Building the scale: public engagement as a scientific need''), and conclude the article with the outlook, summary and conclusion (Sections \ref{sec-outlook} and \ref{sec-sum}).
\end{justify}

\section{Foundations of the CREDO methodology}
\label{sec-phil}


The CREDO experiment by its very idea of making discoveries and approaching truth in various research areas, touches something one can call the borderline of the current state of our knowledge. As explained below, there are reasons to anticipate that the experiment has the potential to falsify some of the adopted models. In the context of CREDO one can even ask questions about
science itself and its methodology.

As an example one can consider exotic QED processes which are potentially within the observational scope of CREDO. Namely,  pair creation or photon splitting in strong magnetic fields. These predictions of QED can now be tested in the context of cosmic rays physics under extreme astrophysical conditions met in pulsar magnetospheres (with typical magnetic field strengths of $10^{12}$ Gauss, or even $10^{14}$ Gauss in magnetars). In such conditions there are several observational signatures of the two processes \cite{Usov_2002, 1996A&AS..120C.111H}. 
What is more, the photon splitting phenomenon (in exotic scenarios) fits very well in the context of cosmic ray ensembles (CRE) that are potentially within the observational scope of CREDO (as long as the opening angle of the secondary photons would not be too large). Therefore, the CREDO experiment opens up new opportunities to test the well-established QED theory as well as consider more exotic scenarios. The experiment  is potentially capable of changing our view of the basis of science, specifically by providing a new type of means to falsify generally accepted theories in an energy range that has not been yet available for to date observatories. In this way one arrives at the strictly philosophical  question to be addressed and that naturally arises in the context of CREDO -- the issue
of parsimony of the scientific method or Occam's razor.

It has been accepted that the scientific method of explaining and understanding facts should follow the principle of not multiplying entities without necessity. Explaining all new phenomena should be based on what one has already had, namely, based on the theories and models that successfully worked so far, and only if this attempt is found to have failed one may consider rejecting these theories and models. 
Occasionally, the parsimony principle, when understood in a fundamentalistic manner, 
would stand in the way of knowledge. This kind of epistemological attitude might lead to supplementing old models and
theories with new ``epicycles'', rather than encouraging one to humbly admit that some of 
the assumptions being adopted so far need to be rejected as inconsistent with
reality.
A solution to this state of things is to presume 
that the parsimony principle, although an invaluable component of the scientific method,
is not something the scientific method entirely stands on. Sometimes
one has to bypass the strict rule (which in practice is done by the majority of the scientific community).
Accordingly, if the CREDO experiment (or some other experiment) comes up with an anomalous result,
not falling within the framework of accepted models and theories, first
the possibility of an error should be considered, be it in the measurement or interpretation side.
However, one should also not be afraid of pushing the limits and going beyond the realm of modern knowledge by extending or redefining the adopted concepts as well as the language with 
which we describe reality.

The CREDO initiative inspires one to deliberate on falsifiability, another important 
element of the scientific method.
It is often said that scientific statements should be falsifiable --
 refutable by contradicting evidence.
This is not definitely true. For example,  general statements about existence
are most often unfalsifiable, nevertheless they are easy to be verified empirically, which makes the statements quite scientific.
Even though science is not entirely determined by falsifiability, this attribute is very advantageous and often 
characterizes  statements formulated by science, either as models or scientific theories.

Classical electrodynamics provides a good example of a  theory that is falsifiable.
For the purpose of illustration imagine that CREDO (or another experiment
of this kind) comes up with  some evidence for spontaneous photon splitting  (to be in touch with the experimental scope of CREDO pointed out earlier)
 -- a phenomenon there is no place for in the linear Maxwell electrodynamics.
 There are effective terms that could be incorporated to the electromagnetic Lagrangian to mimic at the classical level such effects. To be more specific, some QED vacuum polarization effects such as photon-photon scattering or photon splitting in external fields can be effectively described by classical theory (the Euler-Heisenberg Lagrangian is used in this context), however, these nonlinear phenomena are merely interaction effects arising upon quantization of  linear Maxwell fields in accordance with methods of quantum field theory, therefore already explainable within the current paradigm. The real change in the electromagnetic theory required to incorporate new phenomena such as spontaneous photon splitting that could occur in free space and observed through cosmic ray ensemble evolution, would be to replace the Maxwell Lagrangian by an alternative one resulting in nonlinear equations before quantization, or maybe even to alter the quantization method as such.  
Is that to mean that Maxwell's electrodynamics would just be falsified by an observation of a spontaneous photon splitting?

Stated clearly, it is not that simple -- scientific theories are deeply founded,
routinely withstanding repeated attempts at falsification.
In the first place, one should try to explain a particular observational result within the current theory.
Only if such
attempts turn out absolutely unsuccessful and a growing number of anomalies are
observed at the same time, one would rather conclude that the theory has been falsified -- though the old theory
will probably still survive as a good approximation (as this was the case with Newtonian and relativistic mechanics). 
Thus, falsifying a theory is not an easy task, at least not
possible based merely on a single observation or measurement.
This being said, experiments such as CREDO appear even more valuable, since both for the verification or falsification to be feasible, one needs as many channels as possible
through which the universe is looked at.

\section{CREDO within the cosmic ray landscape}
\label{sec-land}

Although CRE detectable on or around the Earth can be initiated by particles spanning a wide range of energies, and since it is not a priori evident whether the chances of observing a CRE increase with the energy of the primary particle, one should stay open-minded about possible focus concerning the energy regimes of investigation. Here, for clarity, we chose to focus on the cosmic rays of the highest energies known, $E$ > 10$^{18}$ eV, hereafter referred to as ultra-high energy cosmic rays (UHECR), capable of initiating CRE composed of billions of component particles which can propagate unaffected large astrophysical distances, as in case of GeV-TeV photons (see e.g. \cite{Risse2007-zy}), of which an observable fraction may reach the human technosphere.

The surprisingly constant, power-law character of the energy spectrum of cosmic radiation observed by more than ten orders of magnitude with an almost constant exponent of about (-3) could be, in principle, continued on without any obstacles. 
At least until the mid-1960s there was no reason to expect any definite end. Although the sources and mechanisms of acceleration of single elementary particles to the energy of a dozen joules both then and now are unknown (the search is still ongoing). Review of objects in the sky that would potentially come into question, their spatial sizes ($L$) and magnetic fields ($B$) suggests that their capabilities ($E_{max} < ZeBL$) do not reach far beyond the limit of 10$^{20}$ eV. It should be remembered that any acceleration mechanism by its nature would have to be of a statistical character and therefore, when analysing the average or typical parameters of primary cosmic ray particles, it is difficult to exclude the occurrence of large and extremely large fluctuations \cite{Tsallis2003-fs}.

The situation changed substantially after the discovery of the cosmic microwave background (CMB) radiation. If we assume that the primary particles of cosmic radiation are protons, as Greisen \cite{Greisen:1966jv}, Zatsepin and Kuzmin \cite{Zatsepin:1966jv} noticed immediately, a sudden end of the spectrum must be caused by collisions of them with CMB photons and resonant production of the $\Delta^+$ particle. $\Delta^+$ decays instantly again into
a nucleon,
and its energy is on average about 20\% less than before the collision. This process occurs as long as the nucleons have enough energy, which happens at about \mbox{5-6}$\times 10^{19}$ eV. However, particles of cosmic radiation do not have to be protons. Observations suggest  (e.g. \cite{Wibig2005-em}), that heavier nuclei with a higher charge ($Ze$), thus easier to be accelerated, may entirely dominate the highest energy cosmic radiation flux.  For nuclei with energies of about 10$^{20}$ eV, when energies are calculated separately for each nucleon, the GZK mechanism does not work, but this does not mean that the Universe for them remains transparent. In the centre of mass system, when colliding with photons of intergalactic radiation (infrared mostly) they excite to a giant dipole resonance, and then fragment, most preferably emitting few neutrons.  Although fragments retain the same energy per nucleon, as a whole they have correspondingly reduced their total energy.

Searching for the end of the cosmic radiation spectrum, according to what has been  just said, could lead to solving the problem of mass composition of cosmic radiation of the highest energy, and thus bring us closer to identifying its sources (and acceleration mechanisms). However, it unfortunately happens that both the GZK cut-off and the photodisintegration of the nuclei start to work effectively in a similar range of energy, just where today's observations of cosmic radiation end because of the scarce cosmic ray flux. In addition to the statistics, another circumstance is that two giant experiments in operation today which statistically dominate the measurements of giant showers in the highest energy range: 
the Pierre Auger Observatory in Argentina and the Telescope Array in Utah, US, 
while claiming a general agreement of the energy spectra within experimental uncertainties up to 10 EeV, they admit the need for non-linearity to bring the spectra in agreement at the higher energies, and within the range of common declinations. However, the sources of this non-linearity have not been identified, yet \cite{Deligny:2020gzq}. It turns out, then, that the current status of the UHECR observations does not allow definite conclusions
about the exact location and nature of the observed cut-off of the energy spectrum, implying the uncertainty about the cosmic ray composition at the highest energies known.

Either way, the results of the leading UHECR observatories and their conclusions from widely quoted publications (\cite{Abbasi2008-zq,Abbasi2014-ed,Abraham2008-lo,Valino2016-gz}) indicate that the spectrum of cosmic radiation is significantly suppressed at the highest energies, although it is not known 
at which energy, and whether this energy depends e.g. on the kind of the sources.
Such a picture is widely accepted and despite minor scratches, small inconsistencies and doubts it seems that we 
understand it.
There shouldn't be many particles above the cut-off and in fact this is the case.
However, among the aforementioned minor doubts, there are still recorded in the last century in several (or actually almost all) large and significant giant air shower experiments, the cases initiated by particles with estimated energies exceeding $10^{20}$~eV.

The first, historical, Volcano Ranch  event was recorded by Linsley in 1962 with the energy of $10^{20}$~eV, \cite{Linsley1963-vr}. 
In the 1980s the Haverah Park experiment reported a significant increase in the number of showers with energies exceeding $\sim 10^{20}$~eV \cite{1980ApJ...236L..71C,WATSON1991116,Lawrence_1991} .
In the Yakutsk array, the record energy was rated at $1.5\times 10^{20}$~eV \cite{1995ICRC....2..756A}.
The Japanese AGASA experiment published the event from December 3, 1993 which had an energy of $2\times 10^{20}$~eV \cite{doi:10.1063/1.1291490}.
However, a world record was set in the USA in the Fly's Eye experiment: $3.2\times 10^{20}$~eV 
\cite{1995ApJ...441..144B}.
These cases have not been discussed recently in literature, but it seems that they require some explanation. The first, more straightforward explanation is that the experimenters might have been misled about the energy reconstruction of their record event by the imperfectness of the tools available at that time, with particular emphasis on numerical tools. Thankfully, the Monte-Carlo methods developed in the 21$^{\rm st}$ century are certainly more precise and they allow today making more adequate corrections than decades ago, in particular in experimental procedures like localization of shower axis, or estimation of shower energy, whether by collecting fluorescent light or determining the charged particle density distribution on the ground. And this explanation could be enough, provided one does not discuss in detail the statistics of the “above 10$^{20}$~eV” cosmic ray detections of 20$^{\rm th}$ century experiments in contemporary analyses. But, as already mentioned, the two great experiments, the Pierre Auger Observatory and the Telescope Array, 
also reported around 20 UHECR cases of energies above 10$^{20}$~eV observed already with the newest tools and methods (see e.g. Ref. \cite{PhysRevLett.125.121106} where 15 events with energies above 10$^{20}$~eV are mentioned collectively). To list just a few such events in detail we mention the
Pierre Auger Observatory measurements which contain an event with energy $1.4\times 10^{20}$~eV 
\cite{Abreu_2010}
\ – or  $1.3\times 10^{20}$~eV \cite{Aab2015-ae}, and the Telescope Array announcement 
concerning an event of similar energy $1.39\times 10^{20}$~eV 
\cite{Abbasi2014-ed}. 
It is of course clear that if the cosmic ray energy spectrum breaks around \mbox{5-6}$\times 10^{19}$~eV, this is most likely a (gradual) suppression rather than an abrupt cut-off, so there must be a few events exceeding $10^{20}$~eV. However, the current statistics of these events does not allow telling whether they are compatible with the spectrum cut-off or not. 

It is a great achievement that today we can tell that the UHECR spectrum has been quite well measured -- see for example two recent Pierre Auger Observatory papers \cite{PhysRevLett.125.121106,PhysRevD.102.062005}, 
and that a lot of effort and resources are being currently invested into explaining the nature of the observed spectrum suppression (GZK-like versus acceleration limit). Nevertheless the current results are still inconclusive, and it is therefore advisable to continue research in the highest energy regime of cosmic rays, and to try alternative methods whenever possible. 
In other words, the status of the dispute in the UHECR area encourages a closer look at the field and being ready for a major revision or breakthrough in the understanding of physics at the highest energies known. The CREDO initiative with its objectives dedicated to going beyond studying individual cosmic rays and taking under investigation also UHECR products -- cosmic ray ensembles, may provide a precious complementary approach to UHECR studies.

\section{UHECR sources and cosmic ray ensembles}
\label{sec-uhecr}


The Standard Model (SM) of particle physics predicts that if cosmic rays (CR) are primarily composed by protons, there should exist a limit on the
maximum energy of the CR coming from distances farther than 
$\sim 100$ Mpc.
This bound is called the GZK limit and was presented in Refs.~\cite{Greisen:1966jv, Zatsepin:1966jv, Stecker:1968uc}. Despite this, UHECR with energies above the GZK limit have been reported, by experimentalist, from directions where there are no nearby\footnote{The current experimental data suggests an extragalactic origin for UHECRs with energies above the GZK cutoff~\cite{Aab:2017tyv}.}  sources \cite{Zotov2017,Verzi:2017hro}. Therefore, it seems that there is a missing piece in our understanding of the sourcing, nature, and/or propagation of the CR. 

The cosmic ray background has been simulated by means of numerical propagation codes~\cite{Kalashev:2000af,Aloisio:2012wj}, and the results have shown that it is very unlikely that CR with energy greater than GZK limit to be photons~\cite{Aab:2016agp}.

In general, one can distinguish two qualitatively different approaches in unveiling the physics of UHECRs: theoretical models assuming interactions of exotic super-heavy matter (including extra dimensions, Lorentz invariance violation, cosmic strings, existence of new particles etc \cite{Weiler:1982qy,Weiler:1983xx,Aloisio:2015ega,Tyler:2000gt,Domokos:1998ry,Coleman:1997xq,Bhattacharjee:1998qc,Bietenholz:2008ni,Scully:2008jp,Gorham:2012qs,Rubtsov:2013wwa,Tasson:2014dfa,Rubtsov:2016bea,Mohanmurthy:2016ven,Klinkhamer:2008ky,Alcantara:2019sco,Aloisio:2007bh,Supanitsky:2019ayx}) and acceleration scenarios describing processes, in which the particles are accelerated by a particular astrophysical object (shocks in relativistic plasma jets, unipolar induction mechanisms, second-order Fermi acceleration, etc.).  Acceleration scenarios rely on the existence of powerful astrophysical sources with available energy that is sufficient for the energy transfer from these objects to cosmic ray particles. 

In the age of multi-wavelength and multi messenger astronomy, transient astronomical objects are of the great interest for UHECRs emission. There are several classes of astronomical objects which can be prime targets for UHECR observations e.g. gamma-ray bursts (GRBs); supernovae (SN); fast radio bursts (FRBs); various classes of active galactic nuclei (AGN) such as Seyfert galaxies, radio galaxies and blazars; and possible neutrino emitting blazars. In the recent past there are evidences that these objects emit or can emit UHECRs. A 5 millisecond bright fast radio burst of extragalactic origin was detected \citep{2007Sci...318..777L}. It is claimed that the radio galaxies have emitted UHECRs \citep{2008A&A...488..879N}. There is an evidence of hadronic $\gamma-$ray emission from supernova remnants \citep{2008ICRC....2..763M}. Blazars are one of the most prominent sources of UHERCs emission. There are several papers which estimated / predicted neutrino and UHERCs emission from blazars \citep[e.g.,][and references therein]{2012ApJ...749...63M,2018ApJ...854...54R}. 
About 2 decades ago, it was predicted that there were some very bright high energy peaked blazars which would be neutrino loud\citep{2002PhRvD..66l3003N}.
And recently, the IceCube Collaboration found the evidence of neutrino emission from the blazar TXS 0506$+$056 \citep{2018Sci...361..147I} which opened a new window to search for such another blazars too.

\subsection{Supermassive black holes as UHECR sources}

Among the most powerful astrophysical sources, one can highlight supermassive black holes (SMBHs) located in the centers of most galaxies due to their compactness and enormous energy available for the extraction. Below we shall briefly show the capability of SMBHs to power the UHECRs in a given model. It appears that up to $29\%$ of the total energy ($M_{\rm BH} c^2$) of the black hole is rotational energy and available for extraction \citep{Bekenstein:1973:PHYSR4:}. Nowadays, all tests of general relativity indicate that astrophysical black holes can be fully characterized by their masses and spins, while other properties of black holes are hidden inside the event horizon and unavailable to the external observer. For a typical SMBH of $10^9$ solar masses the extractable energy is of the order of $10^{74}$eV, turning SMBHs into the Universe's largest and the most compact energy reservoirs. Therefore, it is 
important to search for the processes, which tap these enormous energy sources in the most efficient way.

Attempts to tap the energy of black holes started in 1969 by Roger Penrose \cite{Penrose:1969:NCRS:}, followed by many authors (see, e.g. references \cite{Bla-Zna:1977:MNRAS:,Wag-Dhu-Dad:1985:APJ:,Par-etal:1986:APJ:,2018MNRAS.478L..89D} and references therein), who used the existence of negative energy states of particles with respect to a stationary observer at infinity. If a particle decays into two fragments near a black hole with one of the fragments attaining negative energy, the other fragment (respecting energy conservation) may escape from the black hole with greater energy than those of the primary particle. The remarkable property of the rotating black hole is the existence of a special region outside the event horizon called the ergosphere, where the energy of a particle relative to infinity can be negative. However, the negative energy states may also appear due to purely electromagnetic interactions between the SMBH and surrounding matter without the need for ergosphere \citep{2019Univ....5..125T,2020ApJ...895...14T}. 

\begin{figure}[H]
  \centering
 \includegraphics[width=0.45\textwidth]{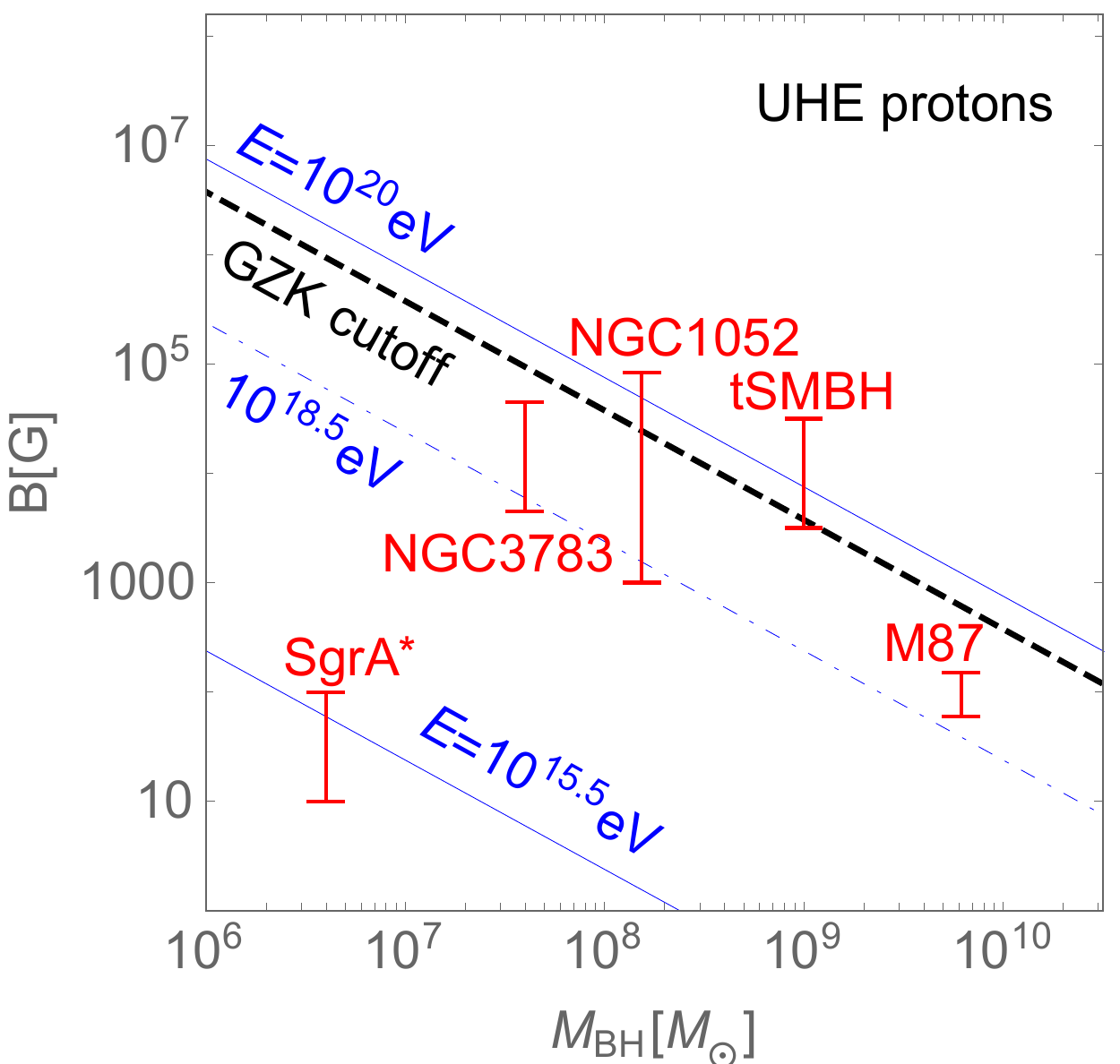}
 \includegraphics[width=0.45\textwidth]{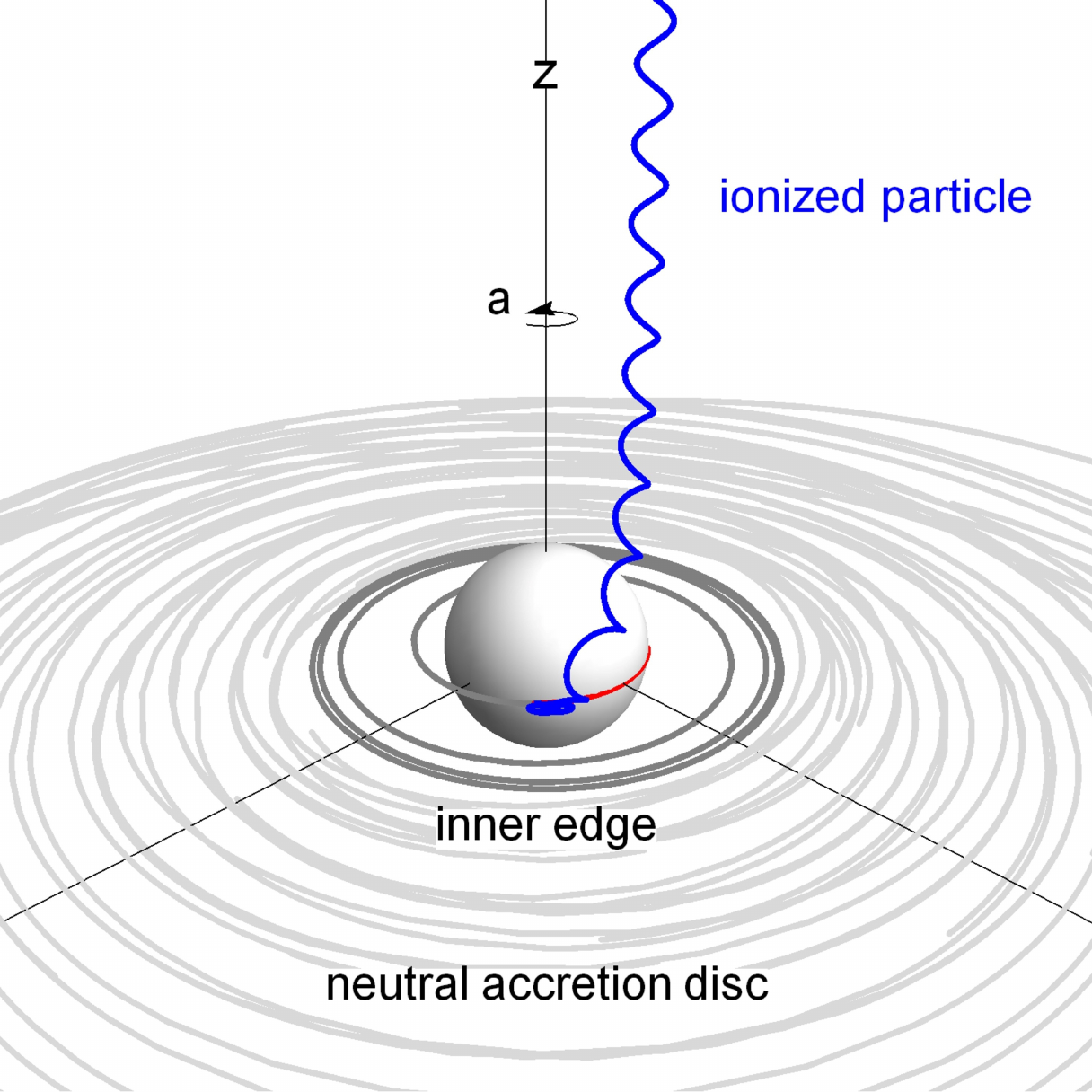}
  \caption{ {\it Left:}
  constraints on the SMBH mass and magnetic field for UHE protons and chosen nearby sources (at the distance $< 20$ Mpc from the Solar system) fitting UHECRs of certain energy. Data is taken from \citep{2016A&A...593A..47B,2012A&A...537A..52E,2012Sci...338..355D,2015ApJ...803...30K,2019ApJ...886...37D,2020MNRAS.495..614P}, tSMBH source corresponds to a typical SMBH of mass $10^{9}M_{\odot}$ and magnetic field $10^{4\pm0.5}$G.  
  {\it Right:} numerical simulation of a decay (ionization) of a freely falling neutral particle (grey thick) into two charged fragments in the vicinity of a rotating black hole immersed in an external magnetic field. Positively charged fragment (blue) is accelerated by the black hole and escapes to infinity along the symmetry axis. Negatively charged fragment (red) ultimately falls into the black hole, extracting its rotational energy \citep[see, details in][]{2020ApJ...895...14T}.}
  \label{figAB}
\end{figure}

Black holes, as any astrophysical object, are immersed into magnetic fields. Rotation of the black hole in an external magnetic field leads to the twisting of magnetic field lines due to the frame-dragging effect. Similar to a classical Faraday unipolar inductor, a black hole rotating in a magnetic field attains non-zero induced electric charge \citep{Wald:1974:PHYSR4:}. This charge has been shown to be non-negligible for any astrophysical black hole and is potentially observationally measurable \citep{2018MNRAS.480.4408Z}. Since the induced charge of the black hole is coupled and proportional to the black hole spin, its discharge is equivalent to the slowing down of the black hole and decreasing the black hole's rotational energy. To demonstrate the process of black hole energy extraction and resulting acceleration of particles to ultra-high energy, one can consider the ionization or decay of a neutral particle into charged fragments in the vicinity of rotating SMBH immersed into an  external magnetic field. If the black hole possesses induced electric charge, the energy of charged fragments after decay of the neutral particle obtains a strong Coulombic contribution in addition to the gravitational negative energy in the ergosphere. Since the induced charge of SMBHs is more likely positive \citep{2018MNRAS.480.4408Z}, these are the protons, which would escape from the black hole with tremendous energy \citep{2020ApJ...895...14T}. This is the ultra-efficient regime of the so-called magnetic Penrose process that can serve as a possible mechanism for acceleration of UHECRs when applied to SMBH candidates. The energy of an escaping UHE particle depends on its charge to mass ratio, strength of external magnetic field and the mass of the black hole. For protons accelerated in this mechanism, the maximum energy is predicted to be
\begin{equation} \label{eq1}
\centering
E_{\rm p} = 1.3 \times 10^{21} {\rm eV} 
~\frac{Z ~ m_{p}}{m} 
~\frac{B}{10^4 {\rm G}} 
~\frac{M}{10^{10} M_{\odot}}.
\end{equation}
In Figure~\ref{figAB} we demonstrate the acceleration of UHE protons resulting from the hydrogen ionization in the SMBH vicinity. Similar results are obtained for the neutron beta decay ($n \rightarrow p^{+} + e^{-} + \bar{\nu}_{\rm e}$). Here we provide verifiable constraints on the SMBH mass and magnetic field as UHECR source. As an example we indicate the few famous  nearby SMBH candidates, which are capable of producing UHE protons of certain energies. It is interesting to note that the  black hole located at the centre of our Galaxy can accelerate particles up to the energies corresponding to the knee of the cosmic ray spectrum.  On the right side of Figure~\ref{figAB} we demonstrate the results of numerical simulation of the magnetic Penrose process  \citep[for details, see][]{2020ApJ...895...14T,Stu-etal:2020:Universe:,2019Univ....5..125T}. Remarkably, the mechanism operates in viable physical conditions for typical SMBHs, without the need for a large acceleration zone or exotic assumptions for black holes. 

\subsection{Axion-like particles as UHECRs}

A different approach to explain the observation of UHECR is avoiding the GZK cutoff via particles beyond the SM. In such a case UHECR must be composed by new particles that must fulfill the following conditions: (i) be stable enough to reach the Earth from cosmological distances; (ii) interact very weakly with the Cosmic Microwave Background and the extra
galactic magnetic fields, in order they do not lose very much energy; (iii)
have a significant flux at their origin; and (iv) interact
strong enough in the near galaxy, with the Sun or Earth magnetic field. 

One of the most popular candidates are axions, which were first proposed by Peccei and Quinn, by introducing an additional global axial symmetry to the
standard model Lagrangian~\cite{Peccei:1977hh, Peccei:1977ur} allowing to solve the strong CP problem dynamically. Additionally, axions were proposed as candidates to avoid GZK cutoff in reference \cite{Mirizzi:2006zy,Csaki:2003ef}. However, it has been shown that it is extremely unlikely that the axion production, together with their conversion to photons by the galactic magnetic field, accounts for UHECR within the present exclusion limits~\cite{gorbunov01_axion_partic_as_ultrah_energ_cosmic_rays}.

A different scenario comes from the consideration of particles with similar features of axions, called Axion-Like Particles (ALPs). The general Lagrangian model involving ALPs read as
\begin{equation}
  \label{ALPlag}
  \Lag_{\textsc{alp}} = \frac{1}{2}\partial_\mu A \ \partial^\mu A - \frac{1}{2}m_A^2 A^2 - \frac{g_{\textsc{alp}}}{4} a  F_{\mu\nu}\tilde{F}^{\mu\nu},
\end{equation}
here $A$ is the ALP field, $m_A$ is the ALP mass, $g_{\textsc{alp}}$ is a
model-dependent coupling between ALPs and photons, $F_{\mu\nu}$ and $\tilde{F}_{\mu\nu}$ are the electromagnetic field strength and its dual, respectively.

The Lagrangian in Eq.~\eqref{ALPlag} provides a decay channel for ALPs to photons (\mbox{$A\to\gamma\gamma$}), which plays a fundamental role in experimental searches. To leading order, the decay width for the above mentioned decay is given by~\cite{Beringer:1900zz,Cadamuro:2011fd}
\begin{equation}
  \label{ALPsDW}
  \Gamma_A\left(A\to\gamma\gamma\right) = \frac{g_{\textsc{alp}}^2 m_A^3}{64 \pi}.
\end{equation}
An important feature of ALPs is that they can be converted into photons (and vice-versa) by means of the Primakoff effect~\cite{Halprin:1966zz}; this effect occurs when a strong external magnetic field exist in the region where the ALPs are propagating. The Primakoff effect can induce an ALP-photon oscillation, similar to the neutrinos flavor oscillation~\cite{Sikivie:1983ip}. This ALP-Photon oscillation changes the polarization of photons traveling in external magnetic fields, providing an additional mechanism in order to detect these pseudoscalar particles~\cite{Maiani:1986md}. The ALP-Photon oscillation  could also produce an apparent dimming of distant sources as well, affecting the luminosity-redshift relation of Ia supernovae, the dispersion of quasar spectra, and the spectrum of the CMB~\cite{Mirizzi:2006zy}.

On the other hand, due to the weak interaction between ALPs and the CMB, this can  travel across the Universe essentially without decaying. The decay length of a particle is given by
\begin{equation}
  \lambda_A = \frac{E_{A}}{\Gamma_A m_{A} },
\end{equation}
where  $E_{A}$ is the energy of ALP.  If we require that the decay length has 
to be at least of the order of magnitude of the observed
universe $R_U$ as considered in Ref.~\cite{gorbunov01_axion_partic_as_ultrah_energ_cosmic_rays},
one finds the following condition

\begin{equation}
  \label{cond}
  R_U \lesssim \lambda_A \equiv \frac{E_{A}}{\Gamma_{A} m_{A} } \quad  \Rightarrow \quad \Gamma_{A} \lesssim \frac{E_{A}}{R_U  m_{A} }.
\end{equation}

The Eq.~\eqref{cond} allows one to establish a restriction on the ALPs coupling, which allow it to reach the Earth from distances beyond the
$R_{\textsc{gzk}}$ radius

\begin{equation}
    \label{CALPs}
    g_{\textsc{alp}} \lesssim \left(  \frac{64 \pi E_{a}}{R_U m_{a}^4}   \right)^{1/2}.
  \end{equation}
This scenario has been studied in
Refs.~\cite{gorbunov01_axion_partic_as_ultrah_energ_cosmic_rays,Csaki:2003ef,Fairbairn:2009zi}, constraining their parameter space according to current experimental data.

\subsection{Dark Matter as a source of UHECR}
\label{s:DM}

Two long-standing problems of contemporary astrophysics can be formulated as: (i) What is the nature of Dark Matter? and (ii) How to explain the existence of cosmic rays with energies greater than $10^{20}$~eV? A plausible hypothesis is that these two mysteries of science can be explained with just one scenario. This is in agreement with  ``{\em Occam's razor}'', which favors the least possible set of solutions to a collection of seemingly independent problems. According to this scenario, Super Heavy Dark Matter (SHDM) may decay or be destroyed via annihilation (see e.g., \citep{CKR99}). This implies a production of super massive -- with rest energies of $E\ge 10^{23}$~eV -- particles in the early Universe, during the inflation phase. Such particles could annihilate or decay presently, leading to the production of jets containing copious amounts of less massive secondaries, possibly or even mainly photons. The energies of these secondaries could easily be of the order of $10^{20}$~eV, the value that seems to be out of reach in the acceleration processes in the potential sources. 

The key prediction of such scenarios in the SHDM group is that the UHECR flux observed on Earth should be dominated by photons (see e.g., \citep{BAYKAL06}). On the other hand, the highest energy events observed by the leading collaborations: Pierre Auger Observatory and Telescope Array, are not considered photon candidates if the present state of the art air shower reconstruction procedures are applied. In fact there are no photon candidates within the energy range where SHDM model should give a photon flux, i.e., for energies above above $10^{18}$~eV, and this non-observation result leads to very stringent upper limits on photon fluxes \citep{Petrera19, AUGER19, TelArray13}. 

However, there are two main concerns about such conclusions. First, the present state of the art analysis does not take into account mechanisms that could potentially lead to a good mimicking of hadronic air showers with the showers induced by photons, e.g., a very efficient splitting of the primary energy into secondary photons/particles and underestimation of photonuclear interaction cross-sections. Second, the present state-of-the-art analysis does not take into account mechanisms that could lead to an efficient screening or cascading of the very-high or ultra-high energy photons on their way to the Earth, e.g., interactions under special cases of Lorentz invariance violation \citep{Klinkhamer:2008ky}, so that the products of such screening or cascading are spread over large areas and thus out of reach of the presently operating observatories, which is interpreted as non-observation of ultra-high energy (UHE) photons. If the first possibility occurs, this would mean that UHE photons might have already been detected and they might be in the data but not properly identified. If the second possibility -- namely that the real properties of cosmic rays and the relevant propagation mechanisms are not well understood -- takes place, then one has little or no chance to detect most of the very high energy photons that travel towards us. Both possibilities obviously question the conclusion about limitations imposed on the SHDM scenarios by the presently accepted upper limits to photon fluxes. Furthermore, such a conclusion can be accepted only if both concerns above are alleviated. A detailed study of this issue is imperative in the forthcoming years. 

An inherent part of the SHDM scenario is the very existence of such DM particles. It is not an overstatement that science has currently no clue of what DM particles are, nor what their properties would be. 

For a long time, the Weakly Interacting Massive Particles (WIMPs) were the most favored candidate. There was a reason for that -- the ``WIMP Miracle''. The number density of the particles that freeze out in the early universe in a certain epoch is set by the balance of the two rates: the production-annihilation rate $n_{X}\langle\sigma v\rangle$ and the Universe's expansion rate $H$ (the Hubble constant at that epoch). Note that $H$ characterizes the temperature of plasma in the universe $T$, so the number density of particles which are in thermal equilibrium can drop exponentially fast with lowering temperature if particles are heavy, $m_{X}c^{2}\gg k_{B}T$. On the other hand, the cross-section depends on the coupling constant (i.e., the type of interaction) and the particle's mass. Next, the mass density of DM in the universe depends on both the density and mass, $\Omega_{X}\propto n_{X}m_{X}$, where $\Omega_{X}$ is the ratio (at present) of the particle-``X'' mass density to the critical density in the universe. Furthermore, the coupling constant describing weak interactions -- the Fermi constant, $G_{F}\approx1.1\times10^{5}$~GeV$^{-2}$ naturally introduces the new mass scale of about $100$~GeV. Finally, if the cross-section is the weak cross-section and the particle's mass is about $m_{X}\sim100$~GeV, then the abundance of ``$X$'' is $\Omega_{X}\propto
\langle\sigma v\rangle^{-1}\sim m^{2}_{X}/g_{X}^{4}$ and $g_{X}\sim0.6$, so $\Omega_{X}\sim0.1$. Thus, ``$X$'' is Dark Matter. This is the ``WIMP miracle'': particle physics independently predicts particles with the right mass density to be Dark Matter. In this scenario, $\langle\sigma v\rangle\simeq3\times10^{-26}\textrm{ cm}^{3}\textrm{s}^{-1}$; since $v\le c$, the gross-section should be $\sigma\ge10^{-36}\textrm{ cm}^{2}$. Numerous ongoing direct-detection DM experiments have reached sensitivity levels that correspond to $\sigma\sim10^{-44}-10^{-45}\textrm{ cm}^{2}$ in the mass range of interest -- from tens of GeV to a few TeV, without a statistically significant and reproducible detection. Thus, the WIMP miracle is at odds with experiment. 

The dismissal of the most favorable WIMP miracle in Dark Matter theory has ignited great interest in alternative scenarios. At present, all bets regarding the non-standard DM scenarios are on the table. There is no deficit of the putative candidates, including the super-heavy dark sector candidates. Here are some examples, as follows. 
(1) Magnetic monopoles \citep{tHooft74, Polyakov74} are topological defects that naturally appear in Grand Unified Theories (GUT) and carry a magnetic charge. The natural mass scale for them is the GUT scale, $\sim 10^{25}$~eV. However, the actual mass of a monopole is not constrained and all mass scales above the one that can be probed in an experiment are considered. Being topological defects, monopoles are copiously produced in a GUT phase  transition, creating a severe over-closing problem $\Omega_{X}\gg1$. This problem is remedied by inflation, which reduces the monopole abundance in the amount enough to not contradict observational and recent cosmological constraints \citep{MedvedevLoeb17}. 
(2) Wimpzillas \citep{KL17} are superheavy DM particles, which mass scale is many orders of magnitude above the conventional WIMP scale, possibly as large as the GUT scale. The WIMP mass cannot exceed hundreds of TeV. Otherwise, heavier WIMPs would over-close the universe, $\Omega_{X}>1$. Therefore, Wimpzillas, like monopoles, are not `thermal relics'. They emerge out of thermal equilibrium right after inflation and their density is not determined by the balance of the production-annihilation rate and the universe expansion rate. 
(3) Planckian-scale particles \citep{GSS16, GPSS18, GPSS19} is a whole class of candidates that appears in a minimal scenario of DM, where only gravitational interactions with the standard model are assumed. There is only one parameter in the scenario -- the particle mass, which ranges from TeV up to the GUT scale. These particles are assumed to be produced by gravitational scattering in the thermal plasma of the Standard Model sector after inflation. For example, the Kaluza-Klein excitation of the graviton in the string theory is one of the realizations of this scenario. The above discussion does not present a complete list of SHDM candidates, of course, but just a few popular examples. There are more theoretically predicted candidates. 

The search for signatures of propagation and/or decay of particle primaries beyond the Standard Model is to be taken seriously and thoroughly. In this regard, CREDO observatory can serve as an indirect search for one of the non-standard Dark Matter candidates: super heavy particles with masses equal or exceeding $10^{23}$~eV. Such particles could be produced in the early Universe around the inflation phase and decay or annihilate presently leading to observable products in the UHE range \citep{Bhatt00}. The interest in SHDM scenario is supported by certain difficulties the electromagnetic theory faces at explaining UHECR acceleration, which is discussed below, in Section \ref{s:EM}.

\subsection{Constraint on electromagnetic acceleration of UHECR}
\label{s:EM}

Electrically charged UHECR traversing a region of size $R$ filled with magnetic field $B$ assumed to be uniform, loses its energy due to synchrotron radiative energy loss \citep{Medvedev03} according to
\begin{equation}
\frac{dE}{dx}=F_{rad}=-\frac{2}{3}\left(\frac{Ze}{Am_{p}c^{2}}\right)^{4}B^{2}E^{2},
\end{equation}
where $A$ and $Z$ are the atomic mass and charge of an accelerated nucleus, $m_{p}$ is the proton mass. The solution to this equation is trivial:
\begin{equation}
E=\left(E_{0}^{-1}+E_{cr}^{-1}\right)^{-1}.
\end{equation}
Thus, for an arbitrarily large initial energy, $E_{0}\to\infty$, the particle energy cannot exceed the `critical energy' threshold:
\begin{equation}
E_{cr}=\frac{3}{2}\left(\frac{Am_{p}c^{2}}{Ze}\right)^{4}\left({B^{2}R}\right)^{-1}
\sim3\times10^{16}\frac{(A/Z)^{4}}{B_{G}^{2}R_{kpc}}~\textrm{eV},
\label{ecr}
\end{equation}
where $B_{G}$ is in Gauss and $R_{kpc}$ is in kiloparsec. 

Furthermore, even if one is to continuously accelerate the particle along its curved path via an inductive electric force $F_{EM}=Ze\,E_{ind}\simeq Ze (v/c)B\sim Ze\,B$, one has the following energy evolution:
\begin{equation}
\frac{dE}{dx}=F_{EM}+F_{rad}=ZeB-\frac{2}{3}\left(\frac{Ze}{Am_{p}c^{2}}\right)^{4}B^{2}E^{2}.
\end{equation}
It has a beautiful solution (assuming the initial energy is small compared to the final energy):
\begin{equation}
E=\sqrt{E_{acc}E_{cr}}\tanh\sqrt{E_{acc}/E_{cr}},
\end{equation}
where 
\begin{equation}
E_{acc}=Ze\, B\, R\sim 9\times10^{23}Z\, B_{G}\,R_{kpc}
\label{eacc}
\end{equation}
is the maximum energy the accelerated particle can reach without losses. This solution has two obvious asymptotic scalings. If $E_{acc}\ll E_{cr}$, i.e., losses are small, one recovers the Hillas constraint \citep{Hillas84} $E\simeq E_{acc}$, and in the opposite limit one has
\begin{equation}
E_{\rm max}\simeq\sqrt{E_{acc}E_{cr}}\sim 10^{20}A^2Z^{-3/2}B^{-1/2}\ ~\textrm{eV}.
\label{f=f}
\end{equation}

\begin{figure}[h]
  \centering
 \includegraphics[width=0.48\textwidth]{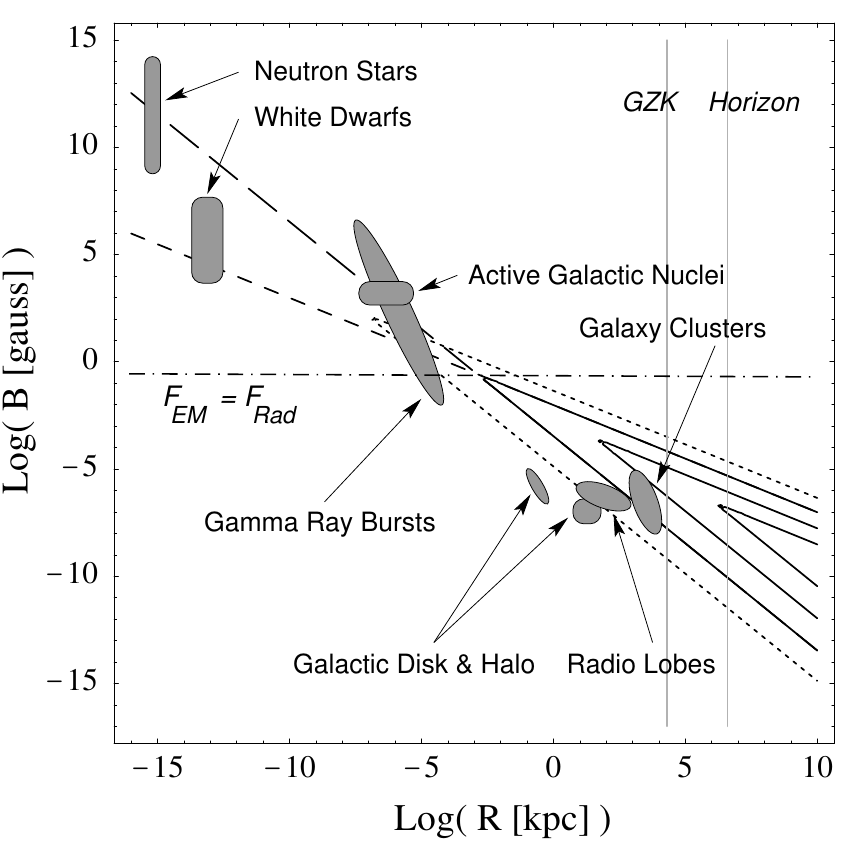}
  \caption{The $B$-$R$ ``diagram of state'' for UHECR acceleration \citep{Medvedev03}. The long-dashed line is the original Hillas relation, Eqn. (\ref{eacc}), for a proton of energy $3\times10^{20}$~eV. The short-dashed and dot-dashed lines represent the radiative cooling constraints given by Eqns. (\ref{ecr}) and (\ref{f=f}), for the same proton energy. Above the dot-dashed line, the force of radiative friction dominates over an electromagnetic force. The solid lines represent the boundaries of the allowed parameter regions for protons with energies $3\times10^{20}$, $10^{22}$, and $3\times10^{23}$~eV, from the outermost to the innermost ``wedge'', respectively. The dotted lines are the same boundaries for an iron nuclei with energy $3\times10^{20}$~eV. Only those astrophysical objects which fall inside the ``wedges'' are, in principle, capable of accelerating the particles to such energies. The grey vertical lines mark two characteristic spatial scales: the GZK attenuation distance ($\sim 20$~Mpc) and the Hubble horizon size ($\sim 4$~Gpc). For GRB sources we took into account that the Lorentz boost changes with radius. }
  \label{figUHECR}
\end{figure}
These results are summarized in Figure \ref{figUHECR}. All the curves, except for two dotted ones, correspond to protons ($A=Z=1$), and the dotted are for iron nuclei ($A=56,\,Z=26$). The details are discussed in the figure caption. Here we make a few important conclusions. First, there is an absolute maximum energy of UHECR accelerated electromagnetically and subject to radiative losses (e.g., in dipolar magnetospheres, galactic and extragalactic shocks and such), which is given by the balance of the acceleration and losses. This puts the upper limit on the magnetic field strength, given by the dot-dashed horizontal line and by Eq. (\ref{f=f}):
\begin{equation}
B\lesssim E_{20}^{-2}A^{4}Z^{-3}\textrm{  Gauss},
\end{equation}
where $E_{20}=E/(10^{20} \textrm{ eV})$. 
Second, there is a corresponding limit on the size of an accelerator. This is the size that corresponds to the ``tip of the wedges'', which can be obtained, for instance, from Equations (\ref{f=f}) and (\ref{eacc}), namely
\begin{equation}
R\gtrsim 6\times10^{-2} E_{20}^{3}A^{-4}Z^{2}\textrm{  pc}.
\end{equation}
This constraint effectively rules out compact accelerators, such as neutron stars, white dwarfs and such. Large objects, such as galactic halos, radio lobes, and galaxy clusters become favorable. Third, electromagnetic acceleration of UHECR beyond 
$\sim3\times10^{22}$~eV is hard because the size of an accelerator becomes comparable to the GZK distance. Furthermore, the acceleration beyond $\sim3\times10^{23}$~eV would require an accelerator of the size of the observable universe, which is questionable. Fourth, our analysis above does not take into account that the source may be moving relativistically with a large Lorentz factor, as in a gamma-ray burst outflow, for example. Accounting for this, relaxes the size and field strength constraints but not very dramatically, as far as UHECR are concerned. Fifth, our analysis excludes special arrangements such as linear accelerators. If one can arrange particle acceleration along a straight path, e.g., strictly along a static straight magnetic field line, then the particle would experience no synchrotron energy loss, regardless of the field strength (still well below the QED, `Schwinger field' strength). In this case, our analysis is inapplicable.

\subsection{SQM objects}

Recently, strange quark matter (SQM) objects (either in the form of stars or planets) have been shown to efficiently convert mechanical energy into hadronic energy when they oscillate \cite{Kutschera_2020}. 
This is possible thanks to the property that the mass density at the edge of SQM objects of $4.7\times 10^{14}\LBgcc$ is the critical density below which SQM is unstable with respect to decay into photons, hadrons, and leptons. Either compression or expansion of the SQM object, such as oscillation induced deformations,  releases energy. Oscillations of SQM objects could be induced in stellar or planetary systems where tidal interactions are ubiquitous. The excitation energy is converted into electromagnetic energy in a short time of  $1\,\LBms$, during a few oscillations. Higher amplitude oscillations result in faster energy release that could lead even to fragmentation or dissolution of SQM objects.  In the context of CREDO, it would be interesting to observe periodic signatures of such events. 

SQM stars and planets are very sensitive to
radial oscillations. 
The amplitude of oscillations is physically determined by the
excitation process in close encounters of SQM objects and
another compact star or black hole. Depending on the closest approach
distance, the energy transferred to the oscillations can reach
$10^{53}\,\LBerg$ when the closest approach
distance is 3 times the star radius. By mode
couplings, monopole (that is, spherically symmetric radial) oscillations also would be excited. 
For violent encounters, the induced oscillations of the radius result in
excitation energy in the
surface layer of every SQM object, equally for stars of pulsar
masses and planetary-mass SQM objects. Fractional amplitudes of radial oscillations $\chi=10^{-6}$ are quite possible. The corresponding
deposited energy to be radiated away for a star of mass $1.4 \LBmsun$ is estimated to be
$6.6\times 10^{36}\,\LBerg$. 
For more violent encounters (such as inspiraling of a tightly bound binary system), one can expect even   
amplitudes $\chi={10}^{-3}$ and the corresponding energy orders of magnitudes higher.
A tightly bound binary system could induce periodic distortions of a SQM companion and associated periodic bursts of intense radiation 
(associated with gravitational radiation). In the evolution of
binary systems with the SQM objects, the energy loss due to
excitations of SQM stars and planets must be accounted for and
this can change significantly the predictions obtained with
unexcited SQM objects (this would be particularly relevant to binary
gravitational wave sources, its exact template calculation should account for 
excitations of SQM objects).

To understand the possible radiation scenarios, one may bring to attention the following estimations taken from \cite{Kutschera_2020}. 
For small SQM objects called planets, the excited energy is deposited in the whole volume of the object. The energy then scales quadratically with the fractional amplitude of radial oscillations $\chi$ --
 for planets $E=7.4\times 10^{48} \cdot \frac{M}{\LBmearth}\chi^2\,\LBerg $. In particular, $E = 7.4 \times 10^{38}\,\LBerg$ for an Earth mass SQM planet undergoing radial deformation of amplitude $\chi = 10^{-5}$.  
As the mass of the considered objects increases, the deposition zone shrinks toward the surface and the energy law changes. In the intermediate region of masses the energy scaling law can be obtained only numerically. 
For masses on the order of solar mass, the energy deposition zone becomes a thin layer and the total energy to be eventually radiated away can again be estimated analytically and it scales cubically with the fractional amplitude of oscillations $\chi$ (the resulting formula is given in \cite{Kutschera_2020}). Furthermore, one also has to take into account a more complicated mass-radius relation for relativistic SQM stars implied by the equilibrium state numerical solution. For  the representative star model of  $M=1.4\,\LBmsun$ the radius is $R=10.3\,\LBkm$ and the energy to be released behaves as $E=6.6\times 10^{54} \chi^3 \LBerg$. For the amplitude $\chi=0.001$ this gives  $E=6.6\times 10^{45}\,\LBerg$ in a $20$m-wide deposition shell.

The process of excitation of quark matter is an irreversible one. The excitation energy will be dissipated eventually into heat
and radiation.  In the intermediate phase the excitation energy 
is being converted into electromagnetic
energy.
The generation rate of radial oscillations of the SQM star can be estimated by
assuming that the timescale of radial oscillations is comparable with that
already known in the model without the excitation energy, which is $T=0.37\times10^{-3}\LBsec$. 
The period is much larger than the timescale for electromagnetic interactions of $10^{-16}\LBsec$.
Thus the excitation energy can be
assumed to be converted into electromagnetic energy inside the SQM star instantaneously. 
The rate of electromagnetic energy generation within a quarter of the oscillation period is $7.2 \times 10^{49} \LBergsec$ for $\chi=10^{-3}$ (and it increases with $\chi$, eg. $6.7 \times 10^{55} \LBergsec$ for $\chi=0.1$). 
Further investigations are required to find  how much of the energy
will be radiated away and how fast this will proceed. A rough conservative estimation made in \cite{Kutschera_2020} shows that
the luminosity for the energy radiated away by the outermost shell of thickness on the order of the photon mean free path would amount to $L=1.3\times10^{34}\LBergsec$ (for $\chi=10^{-3}$) with a corresponding effective temperature of $2.0\times 10^6\,\mathrm{K}$. The total energy of $E=6.6\times10^{45}\,\LBerg$ released in the star would thus sustain radiation for $1.6\times 10^4$ years. However, neutrino cooling will switch on in $10^{-6}\,\LBsec$ and the
star will cool fast. The estimated neutrino cooling time for the volume of the star $1.4\times 10^3\,\LBsec$ is $3.5\times10^8$ times shorter (respectively, $1.5\times10^5\,\LBsec$ for the volume of the initial deposition shell is $3.4\times10^6$ times shorter) than the electromagnetic radiation time.   With the considered excitation mechanism accounted for, the extreme, e.g. $\chi=0.1$ large oscillations, are rather excluded in nature, since the energy of such oscillations would be less than the value of excitation energy generated  during the time of a single oscillation, therefore oscillations would be damped during the first cycle). For lower amplitudes the energy generation time is longer. More accurate calculations would require taking into account the evolution of the temperature of the SQM star. The damping effect on radial oscillations by weak quark processes, although efficient, is not expected to determine the discussed strong interaction phase, although it might dominate the thermalization of excited SQM.

\subsection{Neutron star collapse to third family}

Massive neutron stars may support in their cores exotic states of matter different from hadronic, like deconfined quark matter. These stars are denoted hybrid stars. In the case of a strong first order phase transition inside compact stars, their mass-radius relation presents a gap of unstable configurations between pure hadronic and hybrid stars. The latter ones will populate the so-called ``third branch'', following the hadronic neutron stars and white dwarfs branches. The transition scenarios may correspond to a neutron star configuration lying at the top end of the hadronic branch whose central density is right below the critical value for deconfinement. There are possible ways for the increase of central density of such a neutron star, thus triggering the transition to a third branch configuration. For instance, a fast rotating neutron star may lose energy by dipole emission resulting in a spin down, or a neutron star in a binary system may undergo an accretion-induced spin-up by matter from a companion. The latter case has been recently proposed as an explanation for eccentric orbits of binary systems where the neutron star is able to accrete matter from a circumbinary disk created after a low-mass X-ray episode. Consequently, the neutrino burst which accompanies the deconfinement phase transition in the neutron star interior may trigger a pulsar kick producing the observed eccentric orbit~\cite{Alvarez-Castillo:2019apz}.
A pair of compact stars of about the same mass, each one of them lying in different branches of their mass-radius relation are usually called ``twin stars'', see~\cite{Benic:2014jia,Alvarez-Castillo:2017qki,Blaschke:2019tbh}. The aforementioned dynamical scenarios, where one of the neutron stars collapses into its
twin star, may conserve the total baryon number resulting in a mass defect of less than a tenth of the solar mass for state-of-the-art equations of state, which corresponds to an energy of a few $10^{51}$ ergs~\cite{Alvarez-Castillo:2015dqa}.

It is therefore feasible that the possible deconfinement phase transition in compact stars produce energetic emissions, for instance acting as an engine for Gamma Ray Bursts, with an accompanied characteristic neutrino signal as well as high energy cosmic rays. An analogous situation can be discussed in the framework of neutron star mergers, where both electromagnetic and gravitational radiation are emitted, together with cosmic rays. For the GW170817 event , it has been estimated that the cosmic rays flux is able to support the population of cosmic rays detected at Earth below the ``ankle''~\cite{Rodrigues:2018bjg}. 
%
%
\subsection{UHECR as the spacetime structure probe?} 
\begin{justify}
The CRE hypothesis can be considered a candidate scenario for a number of unexplained cosmic-ray measurements. The examples include Smith \textit{et al}. 1983 \cite{Smith1983-gx} (32 TeV EAS within 5 minutes while only one such event would have been expected) and Fegan \textit{et al}. 1983 \cite{Fegan1983-pn} (simultaneous increase in the cosmic-ray shower rate at two recording stations 250 km apart). On one hand the mentioned measurements were single observations, not seen by other groups, 
but on the other, nobody checked further on a global scale.
We are going to do so within this project -- using the CREDO meta-structure composed of the detectors operated by numerous research groups in all the available energy ranges. This will offer a chance to confirm or question the aforementioned historical reports, potentially leading to the observation of New Physics effects, including probing the spacetime structure. Although the expectations concerning the potential observations of spacetime structure manifestations through effects accumulated over large astrophysical distances are highly uncertain due to missing physics formalism below the Planck scale, as stated e.g. in \cite{Ng2003-ih,Carlip2015-om}, the cumulation of time delays of high energy photons travelling astrophysical distances is thinkable, especially if one considers photons of different energies emitted simultaneously. The cumulative time delay of higher energy photons with respect to the lower energy ones might or might not depend on photon energies -- it is hard to tell due to the missing formalism regarding the proper averaging of spacetime foam fluctuations \cite{Ng2003-ih}. (Due to quantum fluctuations, spacetime, when probed at very small scales, will appear very complicated -- something akin in complexity to a chaotic turbulent froth, which physicist John Wheeler dubbed quantum foam, also known as spacetime foam.) If the cumulative effects of spacetime foam fluctuations are independent of photon energies, then they might be too small to be observed. On the other hand, if they depend on energy, one cannot exclude time delays even as large as of the order of minutes -- comparable to the observations mentioned above. Therefore if, as hoped in CREDO, we can observe CRE composed of high energy photons of different energies, possibly spanning the whole cosmic ray energy spectrum, i.e. from GeV to ZeV, then in any case we will have a new ``spacetime probe'' at hand -- to bring a new input for tuning the existing, or even building the new spacetime models, be it delay evidence or null results that further constrain the available theories. Thus, in any case, we are entitled to expect that within this proposal we will contribute in a novel way to getting closer to understanding the cumulative effects in spacetime foam fluctuations, which would mean a contribution to the foundations of science. While proper averaging methods of cumulative spacetime foam fluctuations are yet to be developed (as stated in \cite{Ng2003-ih}), and despite the uncertainty concerning energy dependence of such effects, not exploring the new research opportunity offered by CREDO would be a methodological mistake. The research in the proposed direction should begin with a wide review of spacetime structure models which predict differences in time delays between the arrival times of high energy photons of different energies. With such a review the experimental efforts dedicated to CRE could be prioritized accordingly so that appropriate detection and monitoring algorithms could be developed, including the alerting mechanisms on a subthreshold level to be analysed in accordance with the multi-messenger astrophysics strategies. The latter direction gives promising perspectives for new advances in astrophysics, reflected also in the new attention of theorists, also those addressing the questions related to space time structure (see e.g. \cite{Carlip2019-so,Wang2019-om,Ng2019-sl}) and private interest in discussions\footnote{P. Homola’s private communication with S. Carlip, Y. J. Ng, and Q. Wang.}.
\end{justify}\par

\section{CRE simulations}
\label{sec-sims}

\begin{figure}[H]
	\centering
		\includegraphics[width=0.8\textwidth]{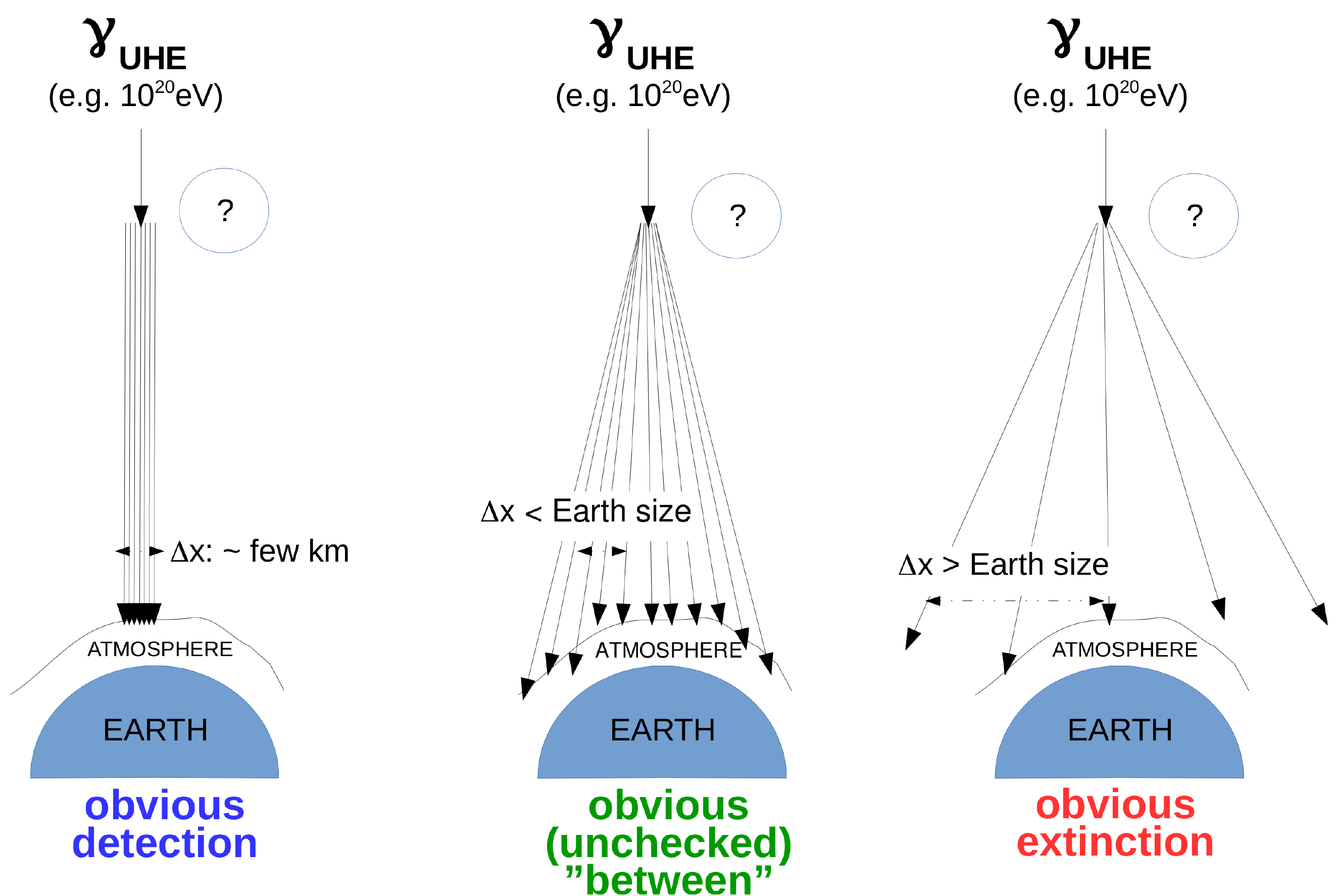}
	\caption{Obviously in experimental reach, although not yet probed, detection of Cosmic Ray Ensembles, using ultra-high energy photons as example primary particles.}
	\label{fig-obvious}
\end{figure}

\begin{justify}
High energy particles that propagate through the Universe unavoidably interact with background radiation, magnetic fields, and matter, initiating cascades of secondary particles which might be observed with available techniques -- like in artificial particle colliders but on a much larger scale. While an interaction of a particle during its propagation through the Cosmos is commonly approximated with extinction, one has to admit that the logically correct question about chances for observing CRE is not if they exist, but when and how we could observe them. This argument is illustrated with Fig. \ref{fig-obvious}. Let us consider a CRE initiated near the Earth, taking as an example the \textit{preshower effect} \cite{Erber1966-ru,McBreen1981-rb,Homola2005-nj,Dhital2018-fn}, i.e. an interaction of an UHE photon and subsequent secondary electrons with the geomagnetic field above the Earth's atmosphere. The word ``preshower'' is meant to describe the result of the initial interaction: ``pre'' emphasizes the location of the interaction vertex above the atmosphere, i.e. before the extensive air showers (EAS) are initiated; ``shower''  addresses the number of particles that emerge instead of one primary UHE photon. Since the distribution of preshower particles (mostly photons) above the atmosphere is confined to a very small region (a fraction of cm$^{2}$) the resultant extensive air shower will have properties very similar to air showers initiated by single primary particles of the energies corresponding to the photon which initiated this preshower. Thus the observation of a preshower-like CRE will require nothing else but a standard infrastructure dedicated to the detection of UHECR, and in Fig. \ref{fig-obvious} (left panel) we name this scenario an ``obvious detection'' example. On the other hand, if a CRE particle distribution is so sparse that only one particle can reach the ideal detector system at a time, there is no technical chance to interpret this individual particle as a component of a CRE. Thus we consider this scenario an obvious technical limitation for CRE observation, in Fig. \ref{fig-obvious}  named ``obvious extinction'' (right panel). Consequently, the focus of any research dedicated to CRE is limited to the scenarios that can result in particle distributions less sparse than the ``obvious extinction'' limit, as also illustrated in Fig. \ref{fig-obvious} with the ``obvious between'' CRE case (middle panel), so far unexplored with a globally coordinated observational effort. 
\end{justify}\par


\par

\begin{figure}[H]
	\centering
		\includegraphics[width=0.9\textwidth]{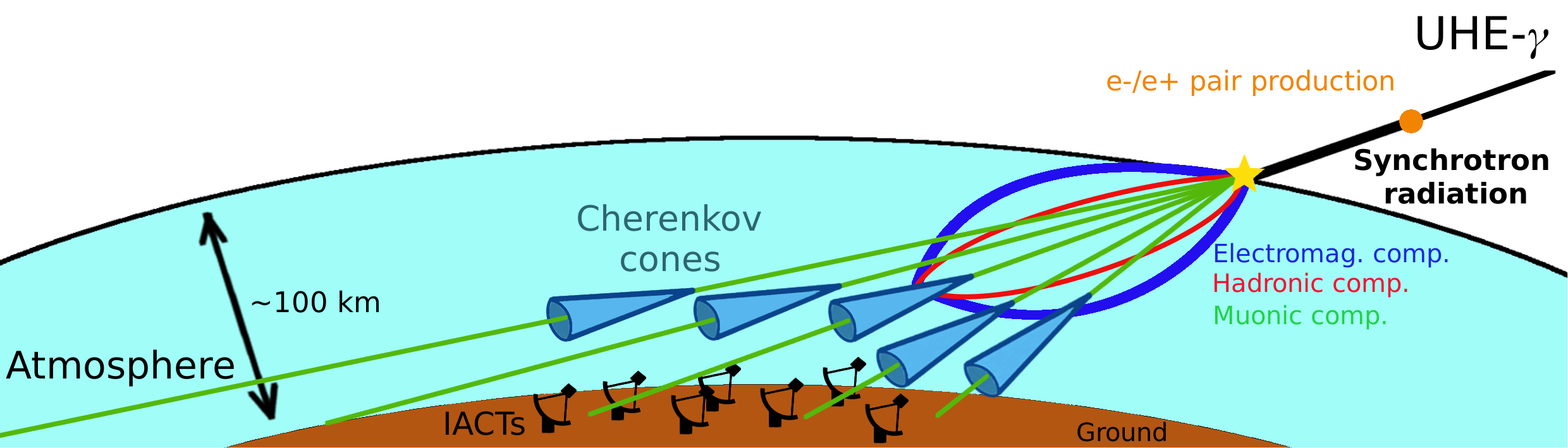}
\caption{An ultra-high energy photon interacting with the transverse component of the geomagnetic field produces an $e^{+}/e^{-}$ pair ${\sim}$1000 km above sea level which emits
bremsstrahlung photons. As such a process can repeat itself for some of these photons, a collection of particles (mainly photons and a few $e^{+}$ and $e^{-}$) reaches the top of the
atmosphere. Consequently, atmospheric air showers are produced and in the case of nearly horizontal showers, only the muonic component reaches the Imaging Atmospheric Cherenkov Telescopes (IACTs) on the ground, which detect the Cherenkov emission of this component.
\cite{Almeida_Cheminant2020-ut}}
	\label{fig-cta}
\end{figure}

\begin{justify}
As far as CRE scenarios are concerned, the ongoing simulation studies include synchrotron radiation of high energy electrons in the presence of planetary, stellar and galactic magnetic fields.

As shown e.g. in Ref. \cite{Risse2007-zy}, in case of the preshower effect occurring due to UHE photon interactions with the geomagnetic field the resultant EAS can hardly be distinguished from those initiated by individual particles based on standard air shower observables like the atmospheric depth of shower maximum development or muon content in the particle distribution on the ground -- unless a large statistics of events is available. Given the stringent UHE photon limits one does not expect many photon- or preshower-induced events to be observed even with the largest air shower detector arrays like the Pierre Auger Observatory or Telescope Array. Instead, one might consider alternative observables and alternative infrastructures, more sensitive to electromagnetic multi-primary EAS origins, e.g. those connected with Cherenkov emission induced by air shower particles and being observed by gamma-ray telescopes. A study in this direction was presented in Ref. \cite{Almeida_Cheminant2020-ut}, where the authors analyze the feasibility of detecting preshower-induced EAS using Monte-Carlo simulations of nearly horizontal air showers for the example of the La Palma site of the Cherenkov Telescope Array (CTA), as illustrated in Fig. \ref{fig-cta}. 
\end{justify}\par

\vspace{\baselineskip}


\par

\begin{justify}
It was demonstrated that there is a realistic chance for identification of preshowers induced by 40 EeV photons coming from an astrophysical point source during 30 hours of observation. This result confirms the Imaging Atmospheric Cherenkov Telescope (IACT) technique could be used to probe physical phenomena not only in the TeV domain, but also in the EeV regime, with a particular connection to the CRE-related physics. Although, as inferred from the upper limits to UHE photons, the rate of expected preshower events for the studied configuration of the observatory is quite low, the gamma/hadron separation obtained by adopting the nearly-horizontal observation mode allows for strong filters to be applied in order to identify such events with a high degree of confidence. Searches for particles with low expected flux using IACTs have been previously performed, as it is the case for the tau neutrino \cite{Ahnen2018-uz} or UHE cosmic rays, as demonstrated in \cite{Neronov2016-hk}. Moreover, multimessenger alerts obtained from other operating observatories may allow fast pointing of the telescopes towards sources suspected to be capable of producing UHE photons, e.g. via interactions between UHE cosmic rays possibly produced by AGNs and the cosmic microwave background, or during gamma-ray bursts, which would significantly increase the chance probability to observe UHE photons and preshower-like CRE. Such potential is well illustrated by the correlation of the arrival direction of a 290 TeV neutrino observed by IceCube with gamma emissions from blazar TXS 0506+056 observed by MAGIC and FERMI-LAT \cite{IceCube_Collaboration2018-yb}. Moreover, a program of observation could be run on catalogs of high energy sources, with an observation time significantly higher than the 30h presented in Ref. \cite{Almeida_Cheminant2020-ut}.
\end{justify}\par


\begin{figure}[H]
	\centering
		\includegraphics[width=0.8\textwidth]{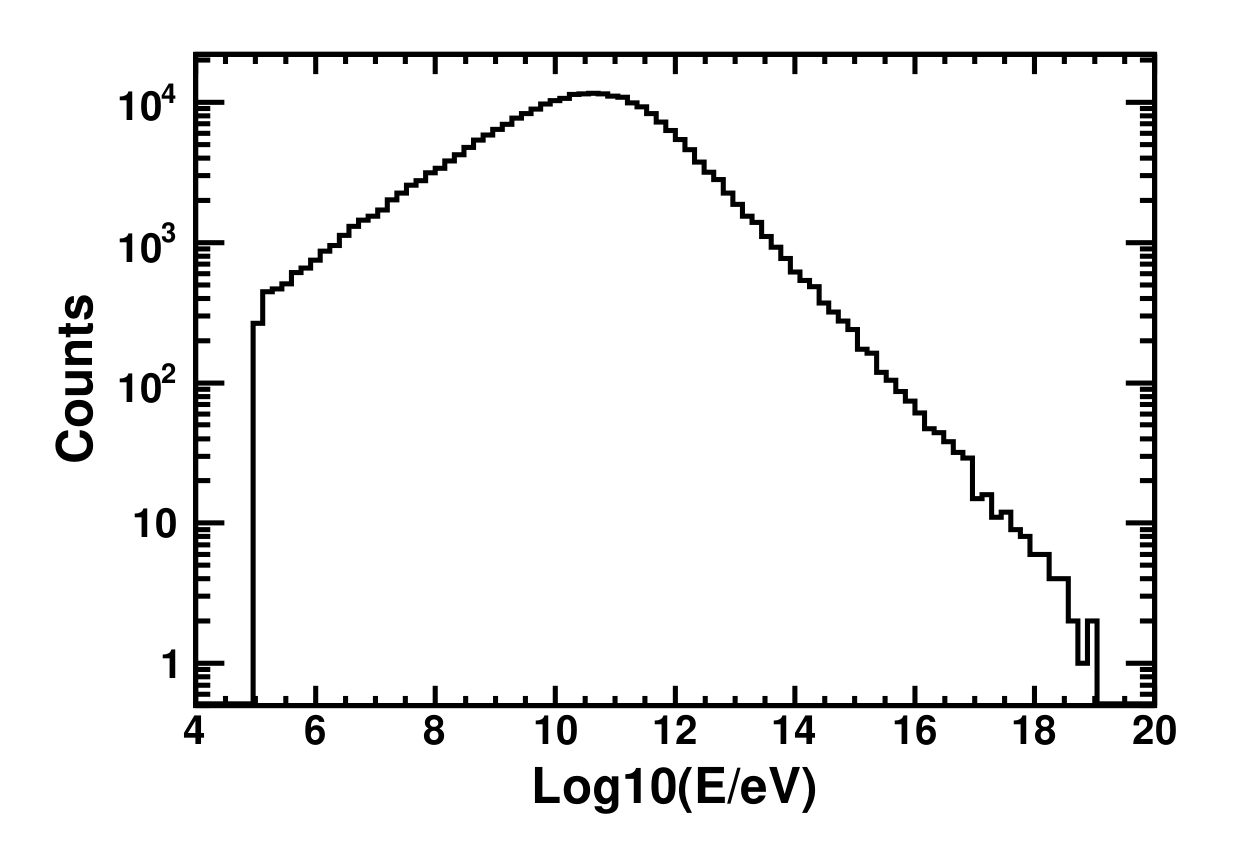}
	\caption{An example energy spectrum of a cosmic ray ensemble generated by an ultra-high energy photon in the vicinity of the Sun, after the interaction with the solar magnetic field (primary photon energy: 10$^{20}$ eV, the heliocentric latitude of ``the closest approach point'': 0$^{o}$, and the impact parameter: 3 R$_\odot$). \cite{Dhital2018-fn}}
	\label{fig-sps}
\end{figure}


\par

\begin{justify}

To generalize the notion ``preshower'', the term ``\textit{super-preshower}'' was introduced (see e.g. \cite{Dhital2018-fn}). 
Super-preshowers (SPS) are cascades of electromagnetic particles, thus a subset of potentially wider CRE family, originated above the Earth’s atmosphere, no matter the initiating process and distance to the Earth. 
Super-preshowers can be classified with respect to their potentially principal observable properties: spread in space and arrival times. For an example, a cascade initiated due to the preshower effect in the geomagnetic field (even at altitudes as high as 10000 km a.s.l.) is expected to have the lateral spread above the atmosphere (100 km a.s.l.) in the order of millimeters, and negligible spread of arrival times \cite{Homola2005-nj,Homola2013-ri}. If the preshower effect was to occur in the vicinity of the Sun, one would expect still negligible spread of arrival times, but the lateral spread then might reach the size of the Earth \cite{Dhital2018-fn} or even larger. The resultant SPS signature is then expected to be composed of even hundreds of thousands of extensive air showers, forming a characteristic, very thin (order of centimeters) and elongated (up to millions of kilometers) pattern.
This example shows that by analyzing the properties of CRE/SPS one might approach attributing a non-trivial physical scenario to the observable event category, provided the underlying uncertainties are properly understood and quantified -- as planned by CREDO. The few calculations performed so far in this direction can still serve only as qualitative indications concerning potential CRE/SPS observables. For instance, in Ref.\cite{Bednarek1999-uv} the lateral spread of an SPS originated nearby the Sun is simulated using a private code and assuming that the SPS is composed only of photons of energies larger than E=10\textsuperscript{17 }eV,\ while from the more detailed calculation done in Ref. \cite{Dhital2018-fn}, obtained with an open source public code PRESHOWER \cite{Homola2005-nj,Homola2013-ri}, one learns that the energy spectrum of SPS particles might be extended over a wide range, down to TeV and lower, as illutrated in Fig. \ref{fig-sps}.
TeV photons would certainly induce air showers that would contain particles, mostly muons, observable on the Earth's surface. While in large cosmic-ray observatories these muons are treated as background, the complete CRE/SPS-oriented research we propose here has to include a proper handling of this ``unwanted muon background'' which can be processed to extract a signal induced by ensembles of low energy air showers arriving simultaneously at the detector -- very clear, unique, and so far untested signature. We make one step further and propose a global analysis of the data from the available detectors to search for extremely spread CRE/SPS events, inaccessible by the largest observatories taken individually. The experimental question which can be addressed in this regard is: ``which CRE/SPS fronts, and in which circumstances, can be detected by a network of devices located on Earth or around it?''. It is the question about the SPS/CRE particle density on the top of the atmosphere (or speaking more general: on top and within the technosphere) and the particle density, or any other observable information, like e.g. Cherenkov light, related to the resultant air shower ensembles. 

Another type of a CRE scenario, where electromagnetic particles play a role, is the propagation of very high energy electrons through intergalactic and galactic magnetic fields. One of the analysis in this direction currently being carried out within the CREDO Collaboration concerns the propagation of electrons of energies between 
10$^{17}$  eV and 10$^{19}$ eV within the Galaxy, using CRPropa \cite{Batista2016-pk} -- a Monte Carlo simulation of cosmic ray propagation and the state-of-the art modeling of the galactic magnetic fields to quantify the chances of observing a CRE on Earth. The intermediate results indicate qualitatively that one might have a chance of observing a CRE originating from synchrotron radiation occurring 
within the Galaxy or maybe event at some other extragalactic sources.
Quantification of the observational chances is still on the way, but already the qualitative study conveys a very important message: even ``conventional''  and abundant electromagnetic processes like synchrotron radiation in galactic magnetic fields are expected to generate CRE reaching the Earth.

At the end of this section it is worth emphasizing that the experimental strategies dedicated to CRE do not need to rely on specific theoretical scenarios -- one might also ``fish'' for clearly non-random cosmic-ray global footprints. This approach is justified by the ultra-high energy physics uncertainties we are aware of -- large enough to admit that we might not be able to imagine \textit{all} the possible physics scenarios predicting a CRE signal that would be observable on Earth. It is therefore sensible ``just'' to explore \textit{fully} the potential of the infrastructure we have at hand and go fishing for signatures we are not able to predict, but which we can distinguish from the diffuse (random) cosmic-ray background. 
\end{justify}\par

\section{CREDO detectors: cloud of clouds}
\label{sec-detectors}

\begin{figure}[H]
	\centering
		\includegraphics[width=0.8\textwidth]{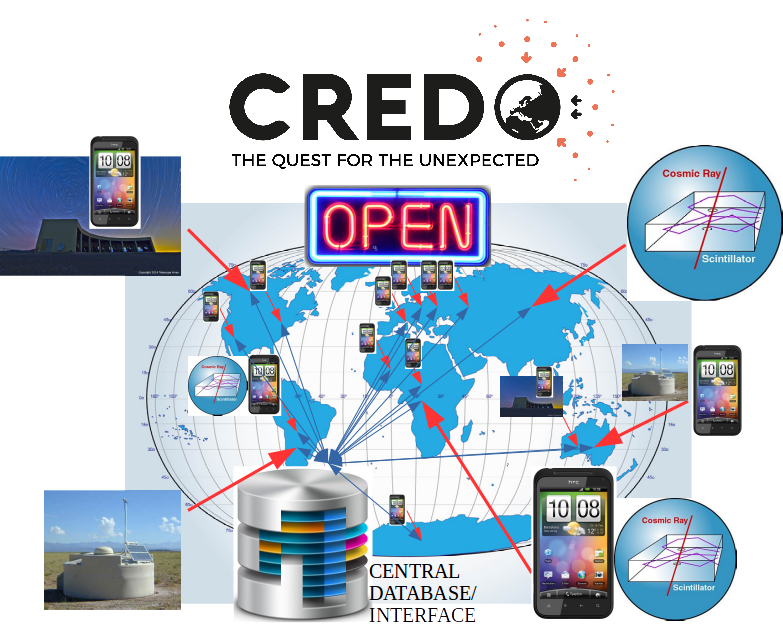}
	\caption{The concept of the Cosmic Ray Extremely Distributed Observatory (CREDO): open on two ends (data upload and access), using both professional, dedicated cosmic ray infrastructure and  off-the-shelf detection solutions, including smartphones.}
	\label{fig-credo-cloud}
\end{figure}

\begin{justify}
As explained above, any experimental strategy oriented on observation and investigation on large scale cosmic ray correlations in the form of widely defined CRE require a global approach to cosmic-ray research. Since both extensive air showers and incoherent cosmic rays might originate from a CRE, one realizes that in fact any detectors capable of detecting cosmic ray signals, both on the ground or on satellites, can potentially serve as valuable components of a global data acquisition system. In other words, the chances for CRE observation increase with any single detector or observatory joining the collective observational effort. Within this general concept, schematically depicted in Fig. \ref{fig-credo-cloud}  the role of CREDO can be understood as an umbrella research program that enables a collaborative effort dedicated to CRE with 
the use of existing infrastructure and expertise,
and with openness for designing and building complementary detectors or arrays -- if required and justified by specific CRE models and research plans.   
\end{justify}\par


\par

\begin{justify}


Widening the participation in the CREDO program increases its scientific potential. The CREDO program benefits all participants by being as open as possible through the following: removing or reducing as many non-scientific barriers that block findability, availability and interoperability of contributing data sets as possible, and facilitating the usage of the systems, taking into account different needs and levels of expertise of the individual participants and institutional stakeholders.
To illustrate technological diversity behind the general CREDO concept in this section, we briefly sketch a landscape of detection techniques used in the state-of-the-art cosmic ray research with particular emphasis on the current situation, needs and interests of the CREDO Collaboration.
\end{justify}\par
\subsection{Cosmic ray detection techniques}
\begin{figure}[H]	
\centering
\includegraphics[width=0.8\textwidth]{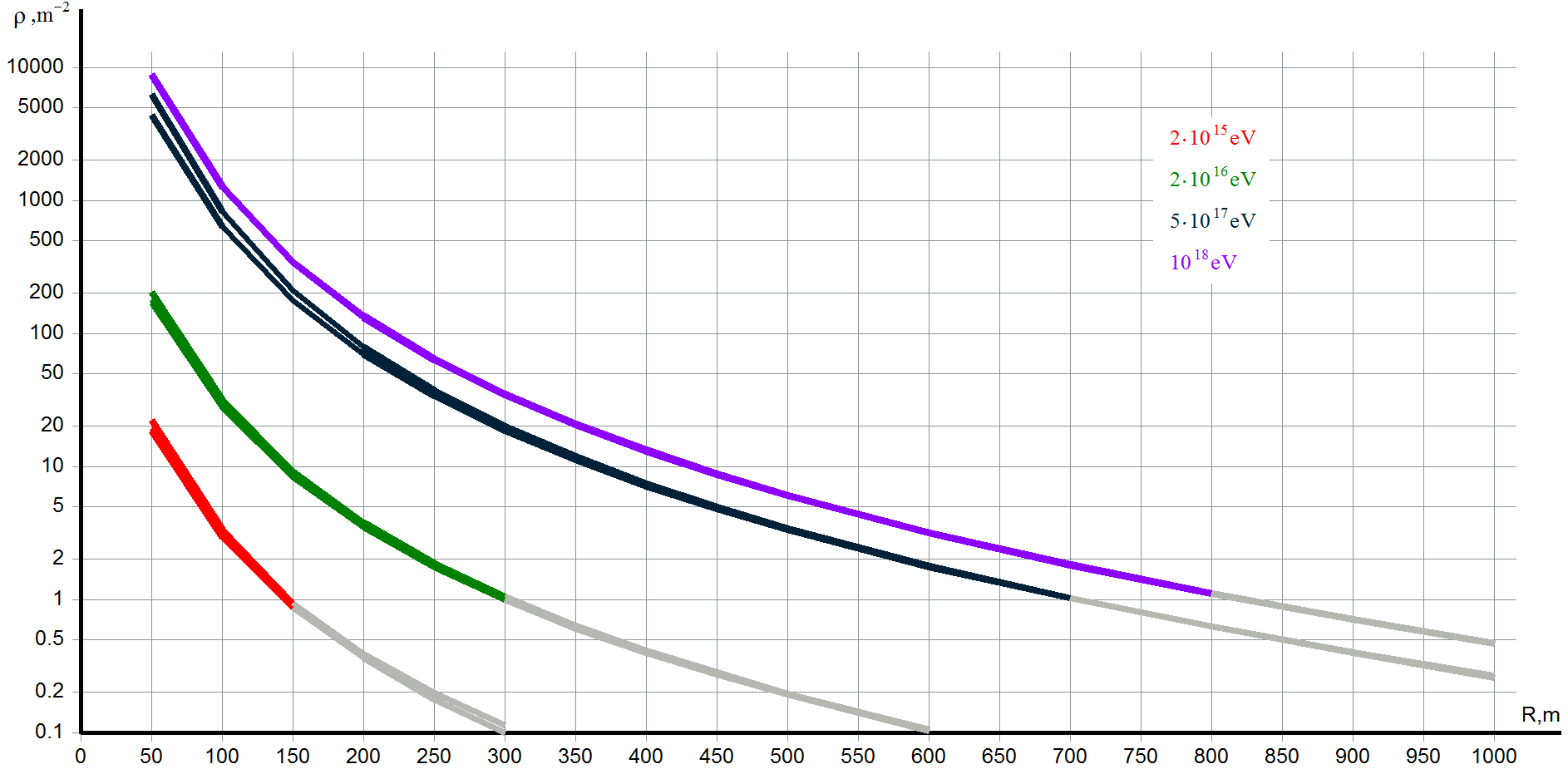}
\caption{Particle density distribution in simulated EAS disk vs. distance from axis for different E\textsubscript{0} \cite{Beisembaev2019-sb}.}

\label{fig:PD_EAS1}
\end{figure}

\begin{justify}
The current CREDO  collaboration with existing experiments and all the future planned detector installations allows for the extension of the widely used EAS (Extensive Air Shower) detection and analysis techniques. EAS detector systems are typically designed as a grid of individual detectors that spreads over a large area. This is due to two reasons: the disk of the EAS at the observation level can be up to a few kilometers in diameter for ultra-high energy events, and the frequency of such events per km\textsuperscript{2} is low so higher  grid area means higher event rate.

The most common technique that is applied as part of  data analysis uses the particle density distribution within the EAS disk. The average distribution of the particle density in the disk varies with the distance from the shower axis in a known way; it additionally depends on the energy of the primary – E\textsubscript{0}. Figure \ref{fig:PD_EAS1} \cite{Beisembaev2019-sb} illustrates the distribution of the particle densities in simulated EAS disks at the different distance from the axis of simulated showers. This is shown for several values of the E\textsubscript{0} of the primary protons. The simulations were done using CORSIKA \cite{Heck1998vt} software package. Note that the particle density distribution is non-linear (approx. inverse quadratic far from the axis). Additionally, each individual detector within the grid provides the time of the hit, e.g. some timing information when the signal in this detector went over a preset threshold. This information is used to determine the EAS arrival direction.
\end{justify}\par


\begin{figure}[H]		
\centering
\includegraphics[width=0.8\textwidth]{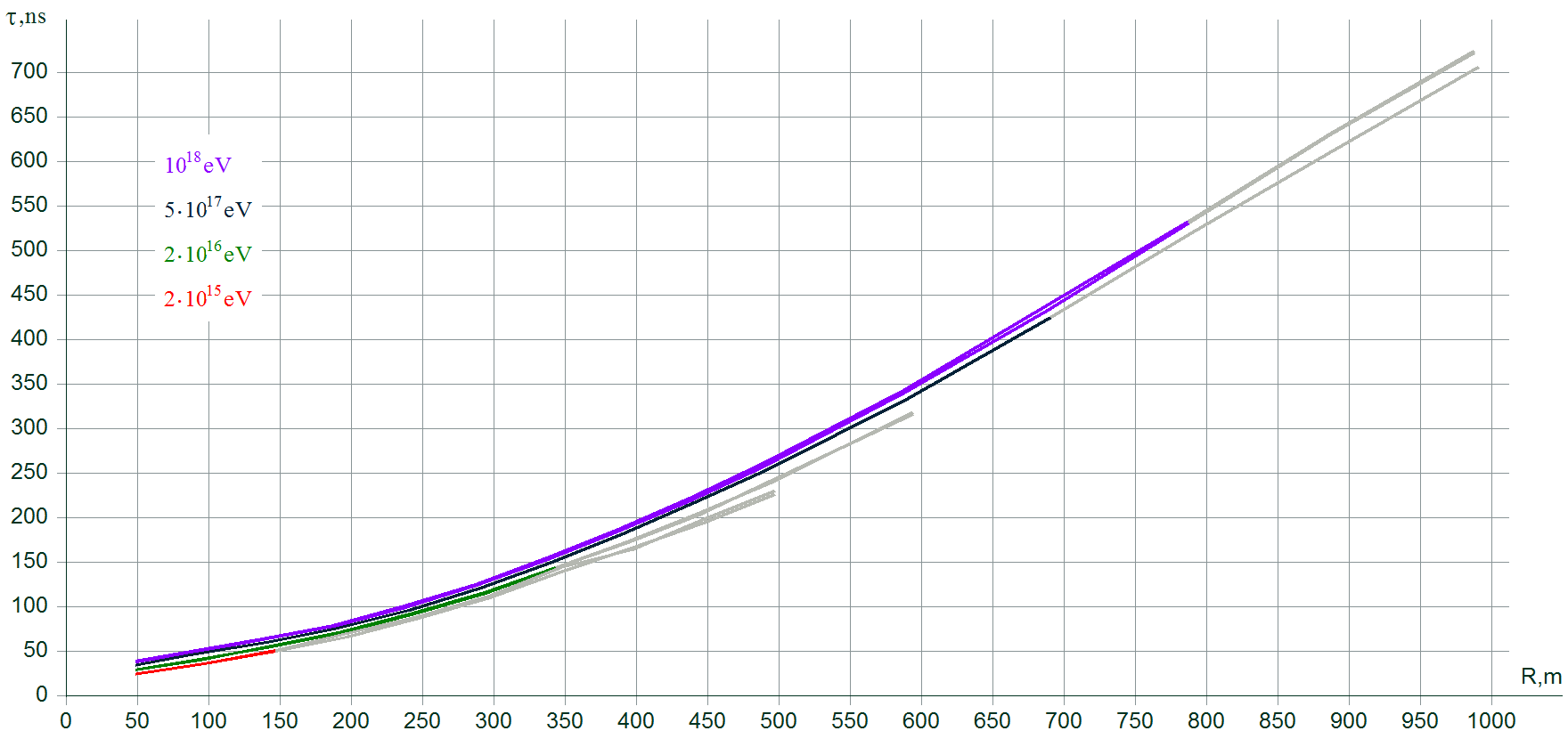}
\caption{Simulated EAS disk width vs. distance from axis for different E\textsubscript{0} \cite{Beisembaev2019-sb}.}
\label{fig:PD_EAS_SIM}
\end{figure}

\begin{justify}
However, there is an unused piece of information here. As the EAS disk passes the detection level, the detectors can measure not only the integral number of detected particles but their distribution over time as well, getting an effective width of the disk \cite{Beznosko2020-be}. This technique was first developed and used at the Horizon-T \cite{Beznosko_2020_PERF} experiment that holds close ties with the CREDO collaboration. This approach uses the width of the EAS disk as measured by different detectors during the event. Figure \ref{fig:PD_EAS_SIM} \cite{Beisembaev2019-sb} illustrates the widths of the same simulated EAS from Figure \ref{fig:PD_EAS1}. The grey lines of the plot show approximated widths at the distances from the axis where particle density is low for reliable determination. The width behaves in a more linear way when compared to particle density, specifically closer to the axis, and is weakly dependent on the primary particle energy E\textsubscript{0}. Using all available information from arrival time , particle density distribution and disk width, one gains advantages such as the ability to do some analysis on the events with axis outside of the detector active area. This ability is due to the fact that the approximate position of the axis position can be reasonably estimated from the disk width information. However, such detectors and specifically electronics are more costly as fast Flash ADC and fast PMT with matching detection medium are required \cite{Beznosko_2017}.
\end{justify}\par


\subsection{Overview of the different EAS detection techniques -- scintillator, water Cherenkov, CMOS/CCD, air fluorescence, radio}

\begin{justify}
Numerous methods are used by existing experiments to detect charged particles from the EAS disk. As we cannot see the particles themselves, the principle is to convert the passage of the particle through the detector into a signal that is easy to process, or observe the result of the particle interaction with the detection medium.
\end{justify}\par
\subsection*{CMOS/CCD}
\addcontentsline{toc}{subsection}{CMOS/CCD}

\begin{figure}[H]
	\centering
		\includegraphics[width=2.39in,height=2.42in]{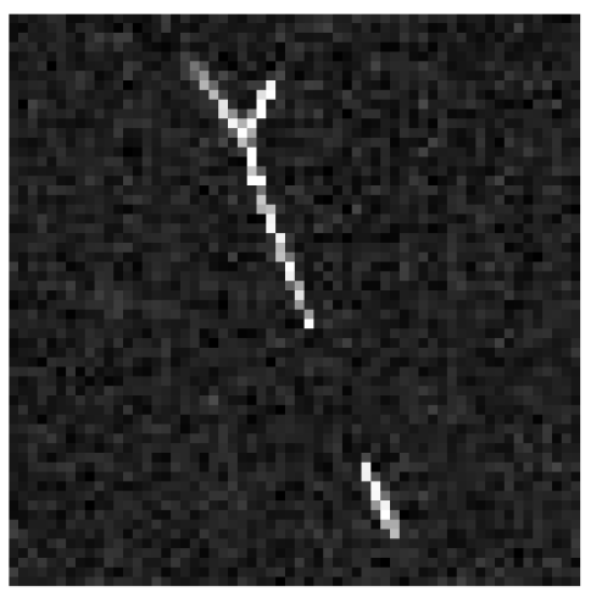}
	\caption{A sample event of a cosmic ray leaving a track on a CCD sensor \cite{Niedzwiecki2019arXiv190901929N}.}
	\label{fig:CR_CCD_TRACK}
\end{figure}

\begin{justify}
The most direct method involves using a silicon-based pixelated strip. The most commonly found detectors of this type are in everyone’s cell phones and photo cameras – they are the CMOS/CCD photosensors (more information is available in \cite{ccd6742594}). On passage through the pixel, charged particles as well as gamma and x-rays affect CMOS/CCD sensors in a similar way as light does. If the camera cap is closed (or cell phone lens is well covered) so that the sensor is in the dark, the pixels affected, or hit, by a particle passage will produce a response as if they were exposed to light. If more than a single pixel is hit, a part of the particle track may be detected as well as shown in Figure \ref{fig:CR_CCD_TRACK}. This method is currently used by the CREDO Collaboration using the cell phone cameras and a special app \cite{Bibrzycki2020-sy,Niedzwiecki2019arXiv190901929N} extendable or portable to a browser application.
\end{justify}\par


\subsection*{Water-based Cherenkov detector}
\addcontentsline{toc}{subsection}{Water-based Cherenkov detector}

\begin{justify}
Another detection method that is based on observing the results of particle-matter interaction relies on the detection of the Cherenkov light – an electromagnetic shock wave from the charged particle passing the transparent medium (water, air, glass, plastic) faster than the speed of light inside that medium (read more on Cherenkov radiation in \cite{Watson2011TheDO}). The shock wave consists of thousands of photons per 1 cm of particle path in the medium that can be detected using fast photodetectors with internal signal amplification, such as PMT (Photo Multiplier Tube).

The design of the water-based Cherenkov detector is very simple – an insulated volume of water and at least one PMT that is optically connected with this volume. A charged particle creates a cone of Cherenkov light that is detected by a PMT. If the volume is very large and many PMTs are used, the Cherenkov light cone can be imaged and used for measuring particle momentum and direction: this technique is used, for example, by Super-Kamiokande detector (more information please see \cite{Walter2008TheSE}).

The advantages of water-based detectors are the detection speed (on the order of one nanosecond \cite{Beznosko_2017}) and relatively lower cost since the detection medium is water (typically, ultra-pure). However, the Cherenkov signal is relatively weak. Other disadvantages are low to no mobility and the insulation of high voltage electric components like PMT from the water.
\end{justify}\par

\subsection*{Scintillator-based particle detectors}
\addcontentsline{toc}{subsection}{Scintillator-based particle detectors}
\begin{justify}
The scintillators are the materials that produce a so-called scintillation light in response to a charged particle passage in addition to Cherenkov light. The main advantages of scintillators over water or glass as detection medium is that there is about 10-15 times more scintillation light photons than from Cherenkov radiation. Also, most scintillators are solid and the detector is easily movable in most cases and does not require a lot of additional external support, but there are liquid and even water-based scintillators \cite{Bignell_2015}. Specifically, plastic scintillators are lightweight and inert solids that can be used for production of detectors that are suitable for installation at schools. In such cases, PMT cannot be used due to high voltage being a hazard, but there are silicon versions of PMT that work at the safe voltage such as SiPM, MPPC, MRS etc. \cite{duspayev2016distributed} \cite{Beznosko:2009uq}. Thus, scintillator-based detectors are most suited for CREDO long-term outreach and educational goals. The main disadvantages are the cost and a slower response when compared to Cherenkov light \cite{beznosko2017horizonMIP} – on the order of 10 ns for plastic and liquid scintillators, on the order of 100 ns for inorganic.
\end{justify}\par

\subsection*{Air fluorescence detectors}
\addcontentsline{toc}{subsection}{Air fluorescence detectors}
\begin{justify}
As the EAS disk travels through the air, the charged particles that comprise it ionize the air molecules along the entire EAS path. As the electrons return to their ground state, the energy is released as light. This light can be collected by typically the collection of mirrors with multiple PMTs as light detectors (there are various designs and detectors are used). The main advantage for this method is that with some luck when viewed at the right angle, an actual development of EAS can be captured and accurate estimates of the properties of the parent particle can be made. The main con is that the observations can be done only during Moonless and clear nights. Fluorescent detectors are often combined with scintillator or water-Cherenkov detectors. More information can be found in \cite{TKACHEV2013ICRC...33.1916T}.
\end{justify}\par

\subsection*{Radio signal detectors }
\addcontentsline{toc}{subsection}{Radio signal detectors }
\begin{justify}
As the EAS disk moves in the Earth magnetic field, the positive and negative charges within it experience Lorentz force in opposite directions and radiate in a radio frequency range around 30-80 MHz or so \cite{Aab:2016eeq}. This effect is highest for the EAS closer to the horizon and moving perpendicular to the magnetic lines of Earth. While many experiments are actively trying to use this method, it is a supplement to all other detection methods listed.
\end{justify}\par

\subsection{The CREDO extension proposals}
\begin{justify}
The involvement of non-scientists and outreach is one of the pillars for CREDO collaboration. There are currently proposals to design and produce portable, most likely scintillator-based EAS detectors that could be used at education establishments (schools, universities) as both demonstrators and part of the education process. The designs are all being proposed with the flash ADC capabilities to utilize this expanded set of analysis techniques that involve using advanced EAS timing information and possibly extend the searches for new phenomena within EAS such as ‘unusual’ events described in \cite{beisembaev2019extensive} \cite{beisembaev2019unusual}.
\end{justify}
\begin{justify}

One of such designs is CREDO-Maze, a project that will create a global, unique physical apparatus, which will consist of a network of local (school) measuring stations. 
The concept of the CREDO-Maze array was developed based on the 20 years old Roland Maze Project \cite{gawin,doi:10.1142/S0217751X05030387}. The technology today has developed greatly and the local shower array idea of Linsley \cite{linsley}  can now be implemented much more easily and, critically, much more cheaply. Eventually it is planned to equip high-schools with sets of at least four professional portable cosmic ray detectors connected locally and forming the small school EAS array. 
Feasibility of EAS detection with such a mini-network was demonstrated in Ref. \cite{Karbowiak_2020} where pocket-size (sensitive surface of $ \sim$ ($25 cm^2$)
 affordable (cost $ \sim $ 100 \$ / piece) scintillator detectors \cite{Axani2018-pu} were used. The CREDO-Maze project uses technologically sophisticated measuring equipment in extracurricular activities: detectors of charged relativistic elementary particles will be made of small ($0.02 m^2$) plastic scintillators. The light pulses will be collected by optical fibers shifting the wavelength from ultraviolet to green and then light will be converted into electrical signals by Silicon Photomultipliers. Further electronics will be based on high speed digital circuits and microcontrollers to connect to higher-level servers via the Internet and WiFi links. Prototypes of individual components of the apparatus have been largely developed independently by several institutions. 

One of the important parameters of proposed equipment is the cost. It is easy to build expensive and complicated, 100$\%$  effective professional arrays. We are on the way of building the prototype which cost (including scintillators, SiPMs, trigger electronics, storage and data transmission micro-computer) is below 200 \$  (compared to  3000 \euro{}  per detector for $\mu$Cosmics detector in \cite{Petropoulos_2020}. Prototypes of individual elements of the apparatus have been largely developed independently in several partner academic centres: University of Lodz, National Centre for Nuclear Studies and Institute of Nuclear Physics in Poland, Institute of Experimental and Applied Physics, in Czech Republic, Swinburne University of Technology in Australia. Completion of the whole and its technical adaptation for replication will be one of the interesting tasks of the project. It is an interesting concept to deliver to some of the end-users (schools) kits, which are adapted for this purpose, assembled only in basic, skill-intensive parts. This would allow students in their local project groups to build and assemble from them a fully operational and efficient whole, under the supervision, of course, of the staff of the institutes managing the local project networks. The independent construction of the operating scientific equipment is an additional motivating element and undoubtedly increases the involvement of young people and the general interest of those not participating in the project. These effects were observed in previous attempts to implement similar activities on a smaller scale.

On the other side it should be mentioned that proposed devices are designed and implemented in such a way that, while maintaining high standards, they are as inexpensive as possible. Technologies will be developed to ensure that the measurement kits can be duplicated and distributed to end users as ``self-assembly kits''  with different degrees of sophistication of the finished components. As potential business projects they will be able, together with educational material pledges and software, to provide a ready-made market product. With positive recommendations based on our research results, the potential market, the demand of educational institutions, seems to be quite considerable. The creation of local structures comprising young people involved and organised in research groups (led by teachers/educators) using network communication and based on science centres, as, e.g., higher education institutions, universities, is an important step in the development and institutional activities research performing organisations, including , as well as research funding organisations. The proposed actions open up new areas of innovation in non-formal non-school education. Creating a model system of social communication networks and demonstrating its effectiveness in the proposed field being an element of STEM will allow to plan and create similar networks realized in other areas of education. There are no contraindications for such networks to cover various groups of young people and research centres.\\

Another concept of building a large scale cosmic-ray network is based on engaging the wide community, thanks to the attractive properties and portability of devices such as pixel cameras. A benchmark standard here is Minipix Timepix-EDU, a hybrid semiconductor pixel detector, with a silicone sensor of various thicknesses (e.g. 300, 500 $\mu m$) provided by ADVACAM \cite{advacam}, using the technology developed within the Medipix Collaboration at CERN \cite{medipix}. This detector provides specific characteristics: fast-data acquisition, portable, lightweight, easy to use and easy to place in difficult-to-access environments, and can be operated remotely. Minipix Timepix EDU is a simplified and price effective version of the Minipix Timepix detector. It was designed and created with the purpose to make science accessible to the public (schools, scientific research centers, non-commercial institutions). The simplified version of the software Pixet, with predefined settings, makes it possible to operate the detector without advanced training in data acquisition just by connecting the device to the USB port of the PC.
\end{justify}
\begin{figure}[htpb]
    \centering
    \includegraphics[width=0.8\textwidth]{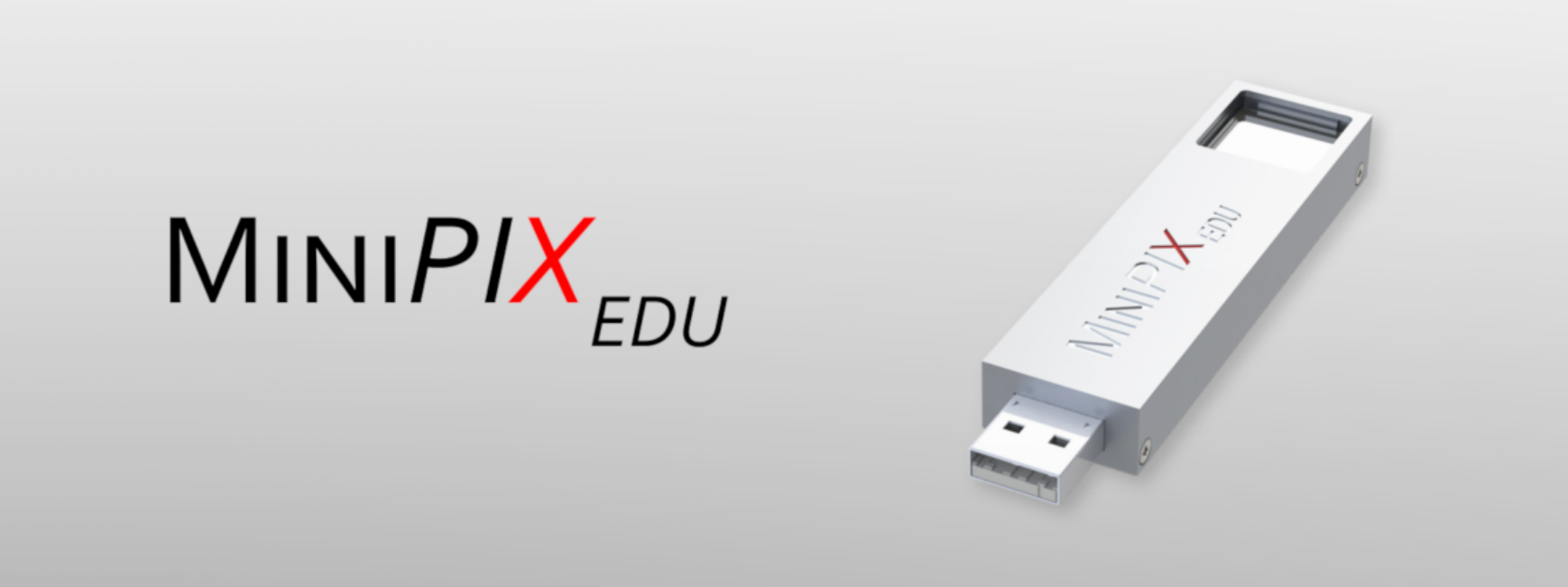}
    \caption{Illustration of the Minipix Timepix EDU detector dedicated for educational purposes, www.advacam.com, accessed on 22.09.2020.}
    \label{fig:BannerWEB_EDU}
\end{figure}
\begin{justify}
The ASIC read–out chip contains a matrix of 256 by 256 pixels (total 65,536 independent channels) and an active sensor area of 14 mm by 14 mm, where one pixel corresponds to 55 $\mu m$, see Fig.~\ref{fig:BannerWEB_EDU}. 
The dark noise signal enables measurements of low–Linear Energy Transfer (LET) particles with high precision. Given to the per–pixel calibration, an adjustable threshold is made in each pixel. This allows a detection efficiency close to 100\% for heavy charged particles, making these detectors unique and suitable for the detection of cosmic rays.
\end{justify}
\subsection*{Online track visualization and processing software}
\begin{justify}
The software package Pixet Pro (D. Turecek, J. Jakubek, 2020, Advacam s.r.o., Prague, Czech Republic) is used to operate the detector and to control the readout, data acquisition and recording. The detectors can be connected to a standard notebook using a USB 2.0 port. The PC connectivity and cross-platform operating system compatibility includes Windows, Linux and Mac OS.
\begin{figure}[h]
    \centering
    \includegraphics[width=0.8\textwidth]{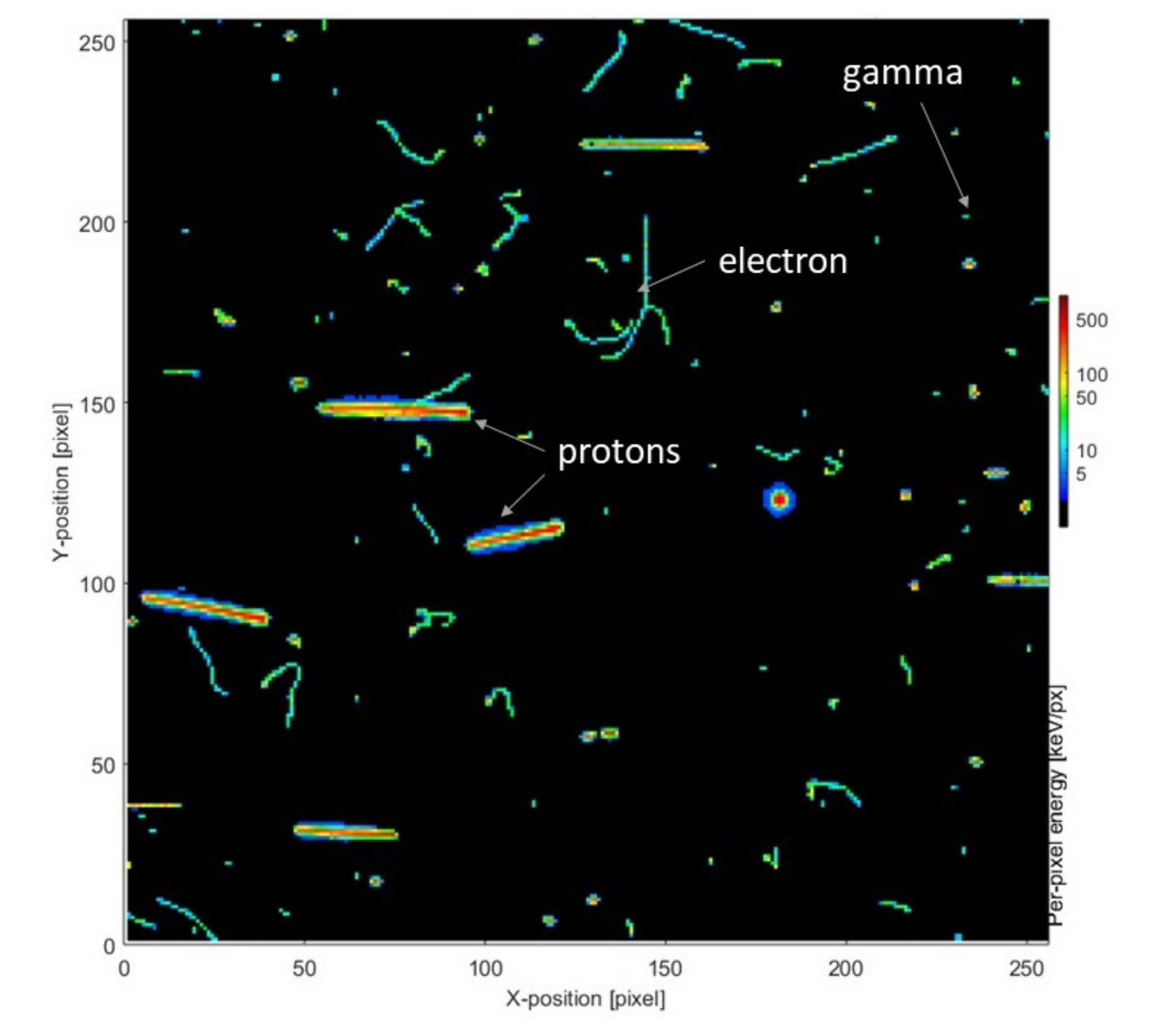}
    \caption{An example of per-pixel energy deposited by various particles in a mixed radiation field measured using a Minipix Timepix detector with Silicone sensor. The detector was operated in ToT-frame mode with an acquisition time of 10 ms. Low-energy, narrow, curly tracks are typical for electrons, high-energy, wide, straight tracks for energetic heavy charged particles such as protons.}
    \label{fig:Minipix_Timepix_detector}
\end{figure}
\\
An example of a mixed field frame can be seen in Figure~\ref{fig:Minipix_Timepix_detector}. Large roundish blobs are created by alpha particles, long strikes by cosmic muons, curving tracks by electrons or small dots by gamma or X-rays. Moreover, rare and exotic events can be observed: Delta electrons, recoiled nuclei, cascade of two or more nuclear transitions, proton tracks. 
According to the frame occupancy the acquisition time can be set. Other parameters such as threshold are already predefined. Further, the data can be processed as dose rates, absorbed dose, fluence maps, energy deposited, LET spectra \cite{GranjaVSAT}.
%
\\\\
A notable role in the implementation of the CRE-oriented detection strategies is also going to be played by professional, medium size detectors giving a high quality signal.


Since 2018, low-background, digital gamma  gamma-ray spectrometer with active shield has been operating in a ground level laboratory in the Department of Nuclear Physical Chemistry, Institute of Nuclear Physics Polish Academy of Sciences (IFJ PAN), Krakow, Poland. The spectrometer is equipped with Broad Energy Germanium detector BE5030 (Canberra, USA), multi-layer passive shield and five large, plastic scintillators (Scionix, Netherlands), playing a role of additional active shield (veto system). Areas of scintillation detectors range from ca. 0.14 up to 0.49 m$^2$. Data acquisition as well as signal processing are performed by means of digital analyzer (so called digitizer) DT5725 (CAEN, Italy) \cite{Gorzkiewicz2019}. 

The main role of both passive and active shielding is reduction of the radiation background  of the germanium detector. However, thanks to digital data acquisition it has become possible to expand the research potential of the constructed spectrometer. Registering and storing data generated by all spectrometer's detectors and using  manifold, off-line data exploration techniques allowed one to initiate continuous monitoring of the cosmic-ray muon flux.

From the CREDO Project point of view, such device may be used as a \textit{reference detector} because scintillators register several dozen of muons per second and thanks to their non-collinear orientation (Fig. 1 in \cite{Gorzkiewicz2019}) it is possible to detect at least 3 muons correlated both geometrically and in time. Preliminary research showed that during single gamma-ray spectrometry measurement, that lasts 426721 s (about 5 days), scintillators registered 329 events of 5-fold  detection coincidences of muons (Fig. \ref{fig:5_fold_coinc}) that could originate from air showers. 

\begin{figure}[htpb!]
    \centering
    \includegraphics[width=.8\textwidth]{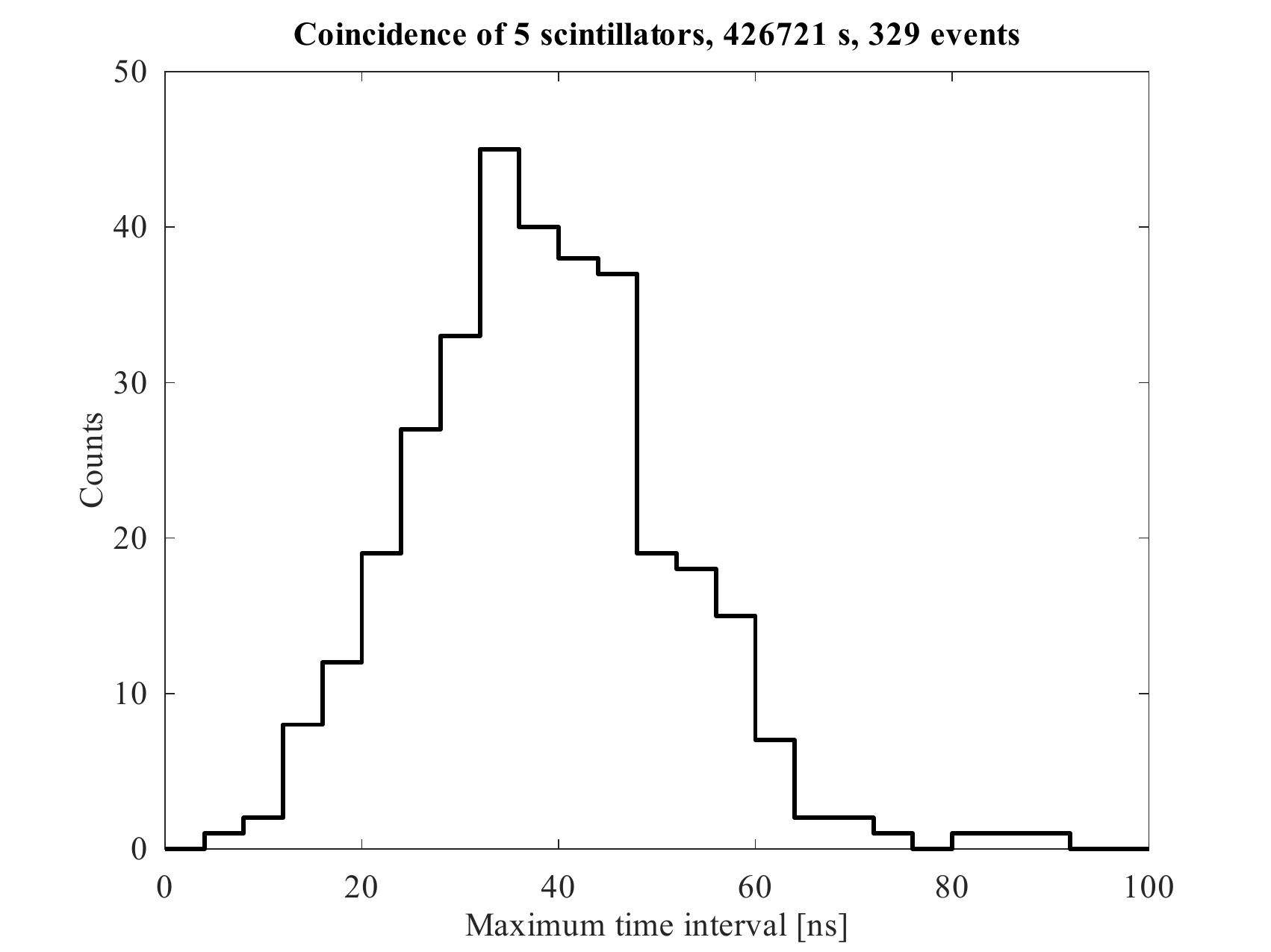}
    \caption{Distribution of maximum time intervals of 329 events of five-fold coincidences registered by spectrometer's active shield during single gamma-ray spectrometry measurement (time of measurement 426721 s).}
    \label{fig:5_fold_coinc}
\end{figure}
Additional advantage of continuous monitoring of cosmic-ray muons flux is the possibility to investigate correlation of  changes in its intensity and Earthquakes that modulate the local geomagnetic field \cite{kovalyov2014}. 

\vspace{\baselineskip}
\begin{justify}
\subsection{Inter-detector communication} 
\end{justify}\par

\begin{justify}
An important aspect of R$\&$D concerning the CRE detection feasibility is communication between detectors constituting an array. It requires attention especially in deployment areas with limited access to internet and/or electricity, where data received by an individual detector cannot be transferred directly to the central acquisition system. Communication might also be critically important in situations which require some preprocessing of the data collected by a subset of the array, before sending a trigger message to the central system. The communication issues concerning the CRE-related applications were addressed in Ref. \cite{smelcerz}. They are the subject of further ongoing investigations and engineering works, below we briefly summarize the current status of these efforts.
 
We focus on a scenario, when designing the network, which assumes that the detectors are mobile devices that can be located in hard-to-reach areas, such as deserts or forests, as in highly urbanized areas, i.e. cities, and even inside blocks (Fig. \ref{fig:scenariusze_label}). Such assumptions force the network to have the following features:

\begin{itemize}
  \item scalability -- it is easy to connect new devices to the network, the network should be self-configuring
\item low energy consumption during data transmission -- an indispensable parameter to ensure support for mobile devices (without access to power from the network)
  \item universality -- the network must operate both in dense urban buildings and in desert or forest areas
  \item wireless -- provides the ability to collect data without the need for expensive infrastructure in the form of cables and the need to use human force (manual data collection from SD cards), in other words, significantly reduces the cost of maintaining the project 
   \item as long range as possible -- guarantees stable transmission without the need to use re-transmitters or uneconomical use of too many gates
\end{itemize}

\begin{figure}[H]
    \centering
    \includegraphics[scale=0.4]{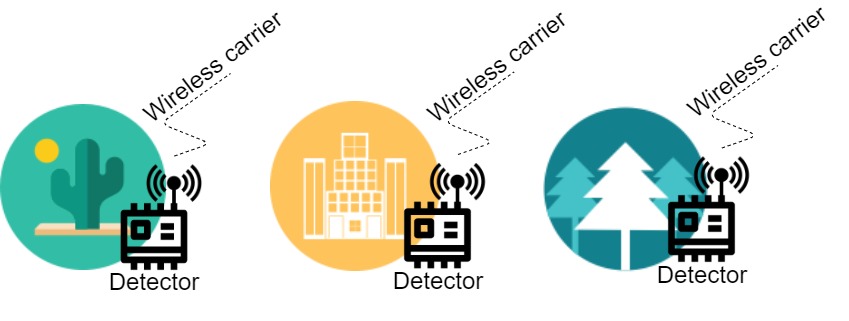} \\
    \caption{Symbolic diagram for the CREDO network scenario.}
    \label{fig:scenariusze_label}
\end{figure}

\subsubsection{Solutions available on the market in wireless networks}

In order to choose the right solution for wireless transmission in the CREDO network, the solutions already existing on the market were compared in terms of parameters important for the CREDO project. The results are presented in Table \ref{tab}. The table completely omits Bluetooth technology due to the very small range (10-50m) \cite{bluetooth}, which does not meet the long range criterion.
\end{justify}
\begin{table}[h]
\centering
\begin{tabular}{ | p{2cm} | p{2cm} | p{2cm} | p{2cm} | p{2cm} | p{2cm} |}
\hline
\multicolumn{1}{|c|}{\textbf{Feature}}& 
\multicolumn{1}{c|}{\textbf{ZigBee\cite{ZigBee}}} & 
\multicolumn{1}{c|}{\textbf{LoRa\cite{LoRa}}}&
\multicolumn{1}{c|}{\textbf{SigFox\cite{SigFox}}} & 
\multicolumn{1}{c|}{\textbf{WiFi\cite{WiFi}}} & 
\multicolumn{1}{c|}{\textbf{GPRS\cite{GPRS}}}\\ \hline
\textbf{Frequency} & 868/915 MHz and 2.4 GHz & 100 MHz to 1.67 GHz & 868/915 MHz & 2.4 GHz & 900–1800 MHz \\ 
\hline
\textbf{Power consumption tx} & 37 mW & 100 mW & 122 mW & 835 mW & 560 mW\\ 
\hline
\textbf{Range} & 100 m & 5 km & 10 km & 100 m & 1–10 km \\
\hline
\textbf{Cons} & Requires infrastructure & Available Gateways on the market are only for 438 and 868MHz & Requires infrastructure similar to GSM (masts and receiving stations) & No Internet connection in non-urbanized areas; high power consumption & No access in non-urbanized areas; high power consumption\\ 
\hline
\textbf{Suitable for battery devices?} & Yes & Yes & Yes & No, too much power consumption & No, too much power consumption \\ 
\hline
\end{tabular}
\caption{Comparison of wireless communication solutions available on the market.} \label{tab}
\end{table}
\begin{justify}

Based on the table above, ZigBee, LoRa and SigFox were selected for testing and deeper analysis. It is worth noting that the ranges given in the table are the best result, often achievable only in an open area, not in built-up areas.
It was decided that all tests will be carried out in built-up areas, considering city centers as the most problematic environment for establishing long-range wireless communication. No retransmitters were used in the tests. The first tests were conducted for the ZigBee network. We managed to achieve a range of about 30m, similar to that presented by \citet{test_zigbee}. Then, tests were carried out for the sigFox network \cite{Discovery_LRWAN1}. Due to the necessity to use ready-made SigFox infrastructure and the still incomplete coverage of the area with this infrastructure, further development work using this standard was abandoned.
Finally, tests were carried out on LoRa modules \cite{Discovery_LRWAN1,Discovery_Spirit1}. We did not manage to achieve satisfactory results (100-300 m). Experiments carried out by \citet{Italy_LoRa_result} in built-up areas in cities that used the actual LoRa network and its ready infrastructure also did not reach the maximum range of 5 km.

After the tests, it became clear to us that in order to obtain good coverage both in the city without the use of repeaters, and in open areas, where there is no network infrastructure. We need to develop our own system. We immediately decided to work at lower frequencies (169 MHz) so as to minimize the attenuation of waves by obstacles \cite{Antennas}.
\end{justify}
\subsubsection{Definition of CREDO Wireless Sensor Detector Network}
\begin{justify}
The CREDO Wireless Sensor Detector Network (CREDO WSDN) is a fusion of the well-known Wireless Sensor Network (WSN) currently used most frequently in solutions for the Internet of Things (IoT).

CREDO WSDN is a wireless network, organized in a star topology, based on radio waves at the frequency of 169MHz, used to transmit information to the Sink collective station, from dedicated mobile cosmic ray detectors. Sink has a COM connection to the PC station, which in turn is connected to the Internet. The internet provides the end user (e.g. a scientist) with access to the collected data. Additionally, additional sensors, such as humidity or temperature sensors, can be connected to the CREDO WSDN to investigate the correlation between events detected by the detectors and other parameters in the area. Fig.\ref{fig:topology} presents the CREDO WSDN topology.
\end{justify}
\begin{figure}[H]
    \centering
    \includegraphics[scale=0.5]{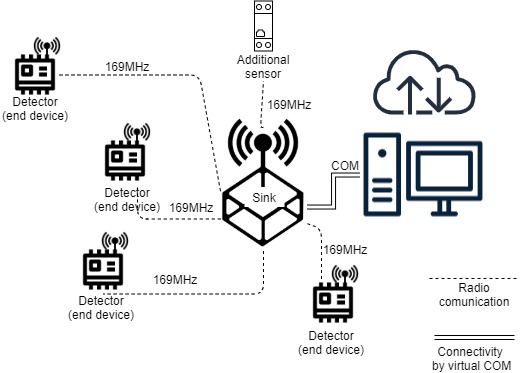} \\
    \caption{CREDO WSDN topology.}
    \label{fig:topology}
\end{figure}
Due to the fact that the network is to be specially adapted to communication on mobile devices (mobile detectors), in the current version of the system communication is only one-way, i.e. the data from the detector is sent to Sink. The radio transmitter will be integrated with the detector itself, not a separate device as it was in the first version of the system. These measures are primarily intended to reduce the power consumption of the detector itself, which is to operate on battery power.\\
The CREDO network consists of several main elements, such as radio communication on 169 MHz, a Sink station, a microcontroller or a power source. The features and role of the most important elements of the entire system are discussed below.

\subsection*{Radio communication}
Radio communication takes place at 169 MHz, unidirectional in a star topology. These features ensure not only excellent coverage in built-up areas or in buildings, but also the optimization of energy consumption -- only the Sink station must remain listening all the time. At the moment, the best result that we have achieved is a range of 8 km in built-up areas.

\subsection*{Transmitter}
We plan to introduce a new version of the transmitter. This year, a new STM32WL chip is to appear on the market -- it is a microcontroller integrated with the radio \cite{STM32WL}. It has excellent sensitivity and transmitter power. At the same time the energy consumption is half less of the previous solution, thanks to the use of a built-in impulse converter. Usually, such a solution significantly worsens the sensitivity of the receiver, but this time it does not have this effect. Additional feature will be that the same system will count pulses caused by particle impacts.

\subsection*{Receiving station (Sink)}
The receiving station in current version already has very good sensitivity \cite{smelcerz} however, as we plan to upgrade the transmitters, there’s a prototype of a new receiving station as well, currently during testing. The new version with a new LNA (Low Noise Amplifiers) amplifier with lower noise and much lower distortion for now increased the range by 4km compared to the previous version (from 4km to 8km).
A new version will be tested as well with several antennas to increase the sensitivity.

\subsection*{Mobile Detectors}
The reason why the network was created was the need to handle the transmission between battery-powered devices, and more specifically mobile cosmic ray detectors. At the moment, the first prototype of the scintillation detector has been developed, the tests are showing promising results, and the device is already adapted to battery power.

\subsection*{Microcontroller}
The microcontroller acts as the brain of the entire system, makes decisions about how to configure the network, when to transmit the collected data, and informs Sink about the transmitter battery status. It is found both in Sink, transmitting stations and in the detector.\\

\subsubsection{CREDO network tests in the field}
During the tests on the first version, it was possible to achieve a satisfactory range of 4 km in built-up areas \cite{smelcerz}. Currently, after changes to the design of the transmitter and the collection station (Sink), the range has increased to 8 km. Also there was significant improvement in range in Elevated Areas (mountains). In the new version of the receiving station, we managed to establish communication on a cloudy day, in the area where the mountain was located. An interesting phenomenon that we were able to observe was the reflection of the wave from the stratosphere on a cloudy day (Fig.~\ref{fig:odbicie}) \cite{Antennas}.

\begin{figure}[H]
    \centering
    \includegraphics{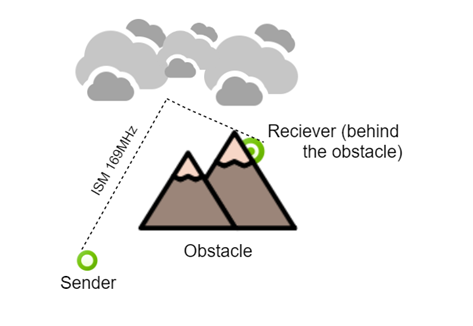} \\
    \caption{Reflection of the radio wave. \cite{Antennas}}
    \label{fig:odbicie}
\end{figure}

\subsubsection{Future work and conclusion}
The CREDO communication system, thanks to its scalability and universality, can be used in many areas of IoT and can be an alternative to existing solutions, where due to the lack of infrastructure or too high costs of their implementation, they cannot be launched.
There are still many challenges ahead of us, we are constantly working on increasing the range. Maintaining security in the network and possibly encrypting data is also a challenge. The phenomenon that we want to study better is the reflection of waves from the stratosphere. We will also be conducting tests with new mobile scintillation detectors in the near future, which will be immediately integrated with transmitting stations.

\subsection{Detection efficiency}

\begin{justify}
Independently of the available cosmic ray infrastructures and expertise that is or might be contributing to the implementation of the CREDO strategies, the novel hardware extensions of the global network of detectors require new studies and tools providing information of detection efficiency on different identification and reconstruction levels: a) individual particles and the corresponding detection rates, b) extensive air showers and the corresponding cosmic-ray fluxes, and finally c) Cosmic Ray Ensembles. The level a) is being addressed elsewhere within this Special Issue \cite{Bibrzycki2020-sy}, considerations of level c) were initiated with Ref. \cite{Verbetsky2020-ea}, and a study dedicated to arrays of portable detectors like CREDO-Maze described above is under preparation.
\end{justify}\par

\section{Data management and analysis}
\label{sec-data}
\subsection{CREDO IT Infrastructure}
From its very beginning the CREDO initiative assumed a necessity for a scalable data acquisition and processing infrastructure.
While most of the infrastructure is supplied by volunteers, in the form of smartphones and other detectors, there still exists a need to create a central repository of detection events. Information about each individual detection event is recorded first by the end device, then it is transferred to the central repository. This repository allows for managing and sharing information among interested parties, it serves as a base for CREDO data analysis. It also fulfills the role of the system’s central point, providing APIs (Application Programming Interface) and information about stored data itself. The CREDO it ecosystem, with some of the implementation details, is depicted in Figure \ref{credo_it_infrastructure}.

\begin{figure}
\centering
\includegraphics[width=\textwidth]{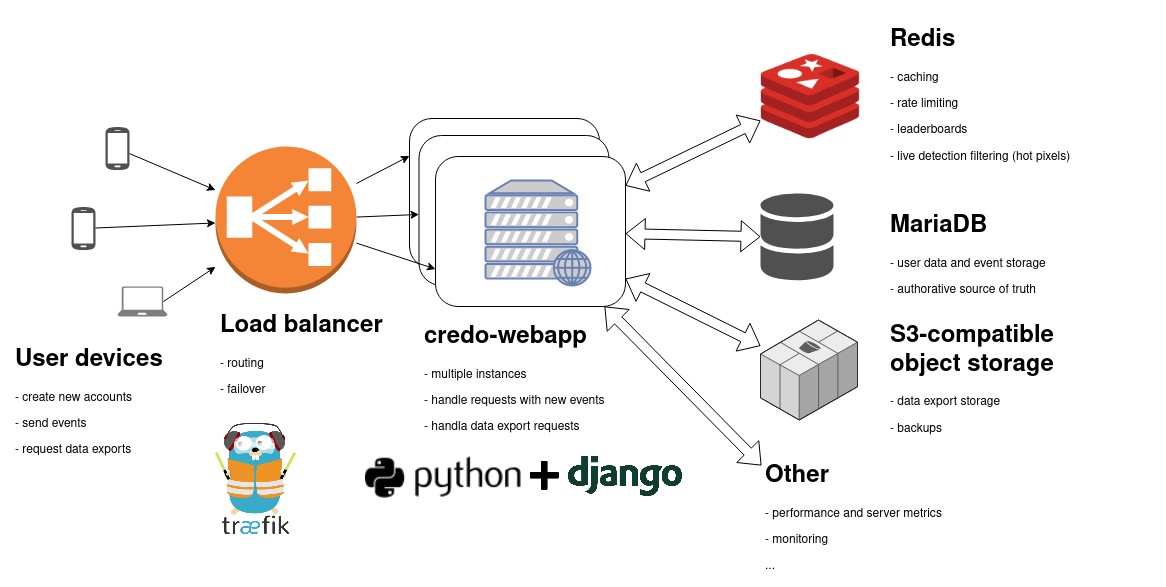}
\caption{Diagram representing the architecture of storage application.} \label{credo_it_infrastructure}
\end{figure}

The CREDO data repository uses a dedicated API, tailored for the project’s needs. The object structure used in the API closely models the information about physical detection events and additionally contains metadata, e.g.\ what detector was used or what was the operating time of the detector. Data storage in CREDO project is provided with respect to the FAIR principles~\cite{wilkinson2016fair}, namely we provide means to Find the data, Access it, enable Interoperability and Reuse of the data. One of the means to achieve this goal is the API, which was implemented as a set of REST services adhering to the OpenAPI standard~\cite{openapi}. OpenAPI is a widely accepted service standard, which includes community-developed guidelines for developing interfaces exposing data and metadata. This standard focuses on data openness, technical accessibility and discoverability, which allow for effortless integration. The stored data is publicly accessible, and so it will be in the future, which should enable the reuse of information and verification of experiments based on the given data.

Central processing for CREDO data storage was implemented as an extensible database, a storage system running in a containerized environment with aim to enhance reliability, portability and security. The main component, responsible for gathering and managing data, was implemented as a Django~\cite{django} web application. User devices communicate with the API exposed by the main application, the data is subjected to basic filtering and then placed in the storage back end. The storage is supplied through the usage of MariaDB~\cite{mariadb} relational database cluster. Redis~\cite{redis}, a key-value in memory object store is used as a temporary cache, implemented in order to speed up queries and data presentation. The external service providing S3 compatible access is used for backups and exporting, prepared beforehand, larger chunks of data.

Due to the nature of observed physics phenomena and the community factor, data stream characteristics are often difficult to predict. Additionally, data analysis is done mainly in an exploratory manner, so the underlying hardware infrastructure needs to be flexible enough to  adapt to requirements of the moment. The presented system has been deployed in a production environment (in this case, virtual machines are hosted in the cloud provided by ACC Cyfronet AGH) and is continuously gathering information about detection events from the CREDO detector network. There are several performance metrics, i. e. CPU usage and request latency, which are constantly monitored thus allowing us to determine if the current hardware configuration delivers required performance. In the case of inadequate resources, the cloud management system is able to autonomously and proactively spawn additional service instances which will allow for load distribution and handling of the data influx. 

It is important to emphasize that the adopted solutions and protocols enable two-end open access: data collected by various detectors can be transmitted to the central system if only their format is kept compatible with the CREDO API structure, and also the access to the data being stored centrally is being granted to everybody upon request including a sensible motivation. The web-based monitor of the CREDO data acquisition system including basic user and detection statistics is available publicly \cite{api_credo} and the technical information regarding the full data access is being provided upon approving of individual requests.   


\subsection{The current data set}
\end{justify}
\begin{justify}
Regardless of the ongoing efforts concerning the FAIR principles, the CREDO data set is continuously available on request for 
individual users if a sensible motivation
for the usage is presented. In addition, all the currently available scripts and tools facilitating data access, selection, and further processing are made freely available in the public CREDO Collaboration repository \cite{credo_github} as a project that can be imported into the PyCharm integrated development environment \cite{Pycharm}.\end{justify}

\begin{justify}
The CREDO data set is divided into three basic categories:
\end{justify}

\begin{enumerate}
	\item Detections -- a set of detections containing detailed information about individual events on all devices;

	\item Pings -- activity logs of devices, including the information about their connections to the database and time of work in the detecting mode;

	\item Mappings -- three collections containing information about users, devices and teams.
\end{enumerate}

\begin{justify}
Most of the data collected by CREDO to date comes from smartphones with the CREDO Detector app, operating on the Android system. The data statistics since the premiere of the app until September 1, 2020 includes:
\end{justify}

\begin{itemize}
	\item 11,150 users (unique accounts) have registered
	\item 15,739 devices were used for particle detection
	\item 4,941,133 candidate detections registered
	\item the total operating time of the devices is over 379,629 days (over 1039 years)
\end{itemize}
The raw data files currently occupy 44 GB (detections: 39.3 GB; pings: 1.6 GB; mappings: 3.1 GB).
Assuming the single CMOS camera sensor has a diagonal in size of $1/3''$ (on average), the total area of all the presently registered devices is 0.56 m$^2$. The daily average value of detections per device calculated from the number of candidate detections registered and the number of total operating time of the devices is about 13 candidate detections registered daily. Currently standard data filtering and clusterization are being performed continuously, and very early results show a perspective for more advanced image processing and classification using machine learning or deep learning techniques to perform more efficient classification of candidate detections.

\begin{justify}
A single detection record is stored in JavaScript Object Notation (JSON) open standard data format and it contains the following information:
\end{justify}

\begin{enumerate}
	\item user -- detection user information: ``team\_id'', ``user\_id'';

	\item location -- geographical coordinates: ``latitude'', ``longitude'';

	\item time -- detection time information: ``timestamp'' (detection unix time in milliseconds), ``time\_received'': reception time in the database;

	\item picture -- detection image information: ``id'' (unique detection identification), ``frame\_content'': image (a fragment of the snapshot, typically containing a margin of 30 pixels around the brightest pixel position -- see below) code in base64, ``height'': resolution ``vertical''  dimension, ``width'': resolution ``horizontal''  dimension;

	\item server side visibility -- ``visible'' (tells whether a detection pass through the server side filters, sensitive e.g. to repeatedly flashing pixels or incompatible versions of the applications sending the data);

	\item brightest pixel position -- ``x'', ``y'' (row and column number of the brightest pixel).
\end{enumerate}

\begin{figure}[H]
	\begin{Center}
		\includegraphics[width=0.8\textwidth]{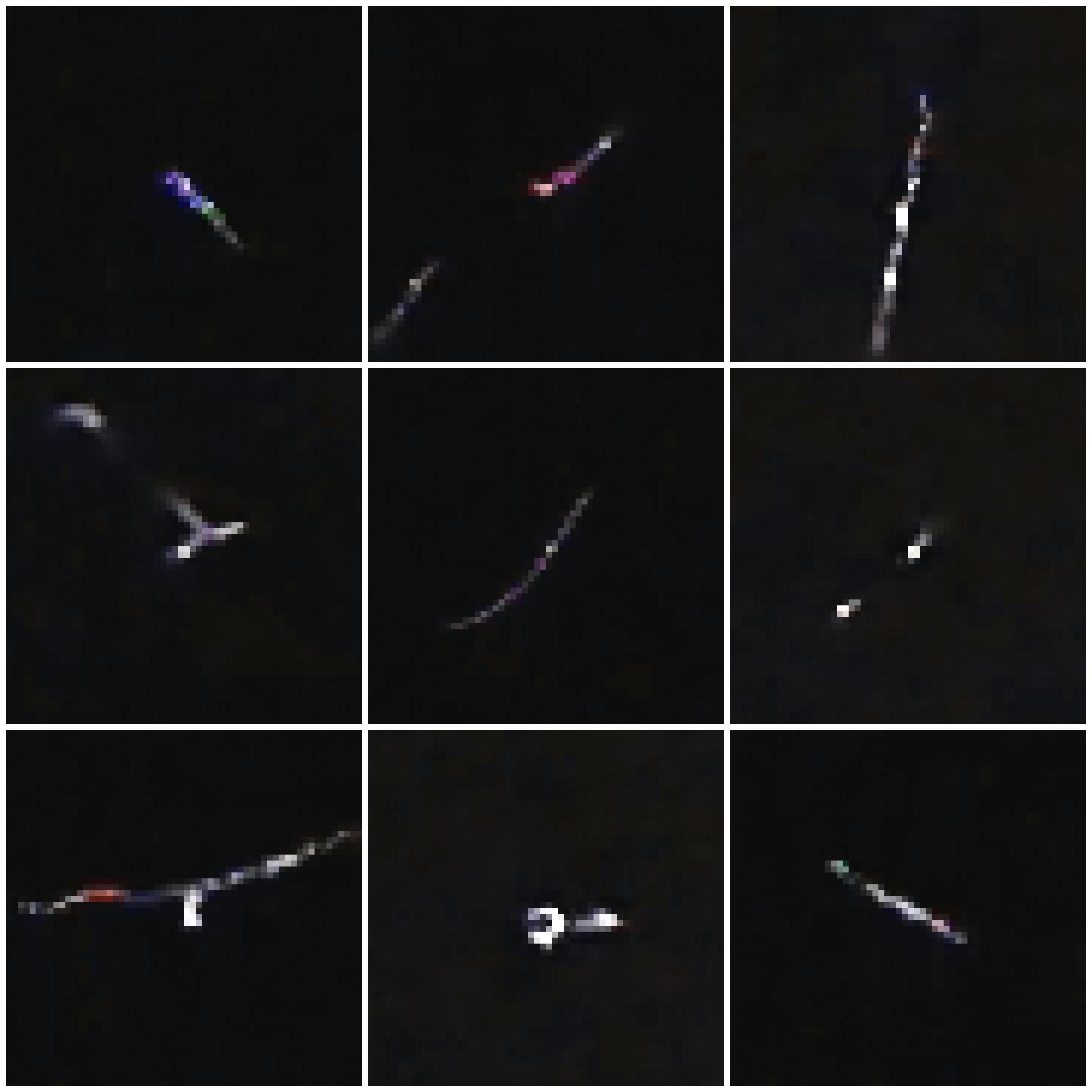}
	\end{Center}
	\caption{Example particle candidate tracks recorded with a smartphone using the CREDO Detector mobile application. [source: the CREDO Collaboration materials and measurements]}
	\label{fig-tracks}
\end{figure}

\begin{justify}
It often happens that a single picture taken by a smartphone in the detection mode contains more than one pixel that fulfill trigger conditions and can be classified as detections. If these pixels are located sufficiently far one from another, i.e. more than the extraction margin explained in p. 4 above, then they are considered separate detections and each of them is assigned an individual detection record. In this way clearly distinguishable particle hits collected in one shot are easily identifiable as detections with the same ``timestamp''. An example set of particle track candidates collected by CREDO Detector is presented in Fig. \ref{fig-tracks}.
\end{justify}


\begin{figure}[H]
\begin{Center}
\includegraphics[width=0.8\textwidth]{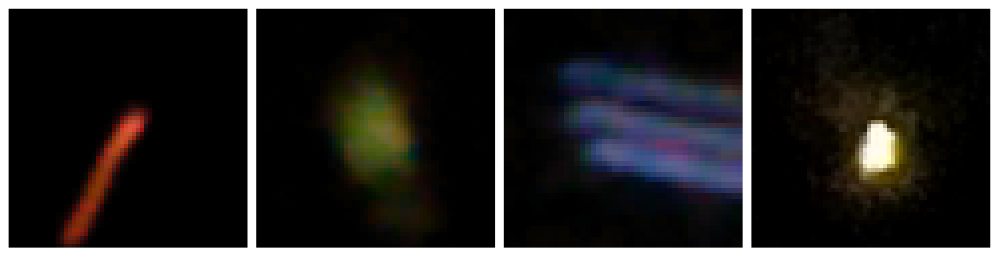}
\end{Center}
\caption{Example artifact tracks recorded with a smartphone using the CREDO Detector mobile application. [source: the CREDO Collaboration materials and measurements]}
\label{fig-artifacts}
\end{figure}

\begin{justify}
The nature of the penetrating radiation measurements made by CMOS sensors assumes identification of pixels that are significantly brighter than the background. The currently used algorithms allow such identification only when the visible light does not reach the sensor, i.e. with the smartphone camera covered tightly. Intentional or unintentional uncovering of the smartphone camera might result in collecting images generated by visible light, sometimes hardly distinguishable from signal excesses induced by penetrating particles (see Fig. \ref{fig-artifacts} for some examples). 
\end{justify}


\begin{justify}
One might apply a specific software filter to remove such a contamination from the data sample being analysed. The internally available preliminary studies in this direction show that the number of bright pixels that compose the image might be a sufficient data quality measure. The rule of thumb currently used within the CREDO Collaboration is based on the data collected by a number of ``trustable'' devices (operated by non-anonymous team members engaged in the application development) and it shows that one does not expect detections of penetrating particles that generate tracks composed of more than 70 pixels with brightness above 70, in the 0-255 scale. 
\end{justify}\par

\begin{justify}
Another type of contamination is the electronic noise which typically gives images composed of individual (or very few) bright pixels only that are being recorded very frequently compared to the expected detection frequency of the cosmic and local radiation. These electronic artifacts can be ``detected'' e.g. if the pixel brightness threshold is improperly set, and such a situation can be identified by monitoring the detection frequency and comparing it to the expected 100$\%$  sensor efficiency for the sensor in use. For an example, a typical CMOS sensor in a smartphone with a surface of 0.2 cm$^2$ can see at most one cosmic background muon in 5 minutes (the expected integrated background muon flux is 1/cm$^2$/minute). Assuming that the local radioactive sources might induce a signal at most $\sim$10 times more often, one does not expect detection rates larger than 2/minute, and this is in fact the case when looking at the statistics collected from the reference (``trustable'') devices. An example of the corresponding filter being used with the CREDO Collaboration requires that detection frequency is less than 10 detections with different timestamps per minute.

Despite the fact that with full access to the raw data sets users are able and welcome to apply their own filters and perform their own analyses, the CREDO Collaboration will periodically release its official data sets and recommendations concerning the data quality which will be based on the appropriate and publically available studies. If it comes to scientific results, the researchers will be expected either to refer to the official data sets, or to describe and justify their own selections.
\end{justify}

\begin{justify}
The examples of the already ongoing analyses concerning the CREDO data set include image classification based on the shapes of particle track candidates \cite{Niedzwiecki2019arXiv190901929N}, monitoring detections collected by individual devices in search for temporal event clustering in 5 minute intervals \cite{Eurek} and muon identification and zenith angle reconstruction based on the lengths of the rectilinear particle tracks \cite{Karbowiak2020-jaa}.
\end{justify}


One of the examples of novel and promising directions of analyses related to the CREDO data is exploitation of the techniques of cyclostationary signal processing and its generalizations
\cite{Napolitano2012}, \cite{Napolitano2019}.
Cyclostationarity is a statistical property of science data generated by the combination/interaction of
periodic and random phenomena.
That is, these data have second- or higher-order statistical functions
that are periodic functions of time. More general models can account
for the presence of multiple, possibly incommensurate, and irregular periodicities 
\cite[Chapters 1,2]{Napolitano2019}.
The observed signals are not periodic, but the hidden periodicity can be
restored by estimating statistical functions from the data. These statistical functions contain
information of the generating mechanism of the data that cannot be extracted starting from the classical
stationary modeling of the observed signals.

The extensions of cyclostationarity can be of interest if time dilation effects have to be accounted for
\cite[Chapter 7]{Napolitano2012}. 
In particular, the effects of constant relative radial speed and/or constant relative
radial acceleration between cosmic source and receiver can be suitably modeled by exploiting the
generalizations of cyclostationarity. General time-warping of the source signal can be
modeled by new and recently proposed signal models 
\cite[Chapter 6]{Napolitano2012}, \cite[Chapters 12-14]{Napolitano2019}.

Source location and parameter estimation problems based on measurements taken at sensors
very far apart separated are very interference tolerant if the source
can be modeled as cyclostationary \cite{Napolitano2020}.
Pulsars have already been successfully modeled as cyclostationary sources \cite{Demorest2011}.

The fraction-of-time distribution of the detected particles can exhibit periodic or
almost-periodic behavior. This behavior can be evidenced
by first-order cyclostationary analysis \cite[Chapter 2]{Napolitano2019}.

\subsection{The data ontology}
The CREDO ambition concerning data quality and access policy is being gradually implemented since the first unique data of scientific value were recorded, stored in the CREDO central system and made public via web services, i.e.\ since the middle of 2018. With the growing data set volume and with the development and improvement of the data mining and interpretation tools, more and more effort was dedicated to maintain the compatibility of the available data with international standards. This effort is presently centered around implementing solutions concerning the data ontology.

The CREDO project is already gathering and integrating detections data originating
from user's devices running the CREDO Detector app, but also from further
sources, and further data, e.g. produced by processing and analysis. One of the
visions of CREDO is also to make this data available for exploitation to the research community,
but also to citizen scientists coming from the general public. This goal is already partly fulfilled by the CREDO data
API described above. However, the current trend in open data publishing is to
make the data a first class web citizen by publishing it in semantic formats on
the Linked Open Data (LOD) \cite{LOD} network%
\footnote{As of July 2020, the LOD network featured at least
1260 datasets from domains such as geography, life sciences, linguistics, and many others, as documented by the LOD Cloud \cite{lod-cloud}.}.
For this, a suitable ontology (also sometimes called linked data vocabulary)
needs to be used, that provides semantic schema for the data. Such ontologies are
usually developed in OWL (Web Ontology Language) \cite{OWL2} and published on
the web. The data is then translated in to the RDF (Resource Description Framework)
\cite{RDF} format using classes and properties defined in the ontology. Larger
datasets are often made accessible via SPARQL endpoints -- web
interfaces by which users and applications may directly query the data using the SPARQL querying language \cite{SPARQL}.

\begin{figure}[H]
\centering
\includegraphics[width=.85\textwidth]{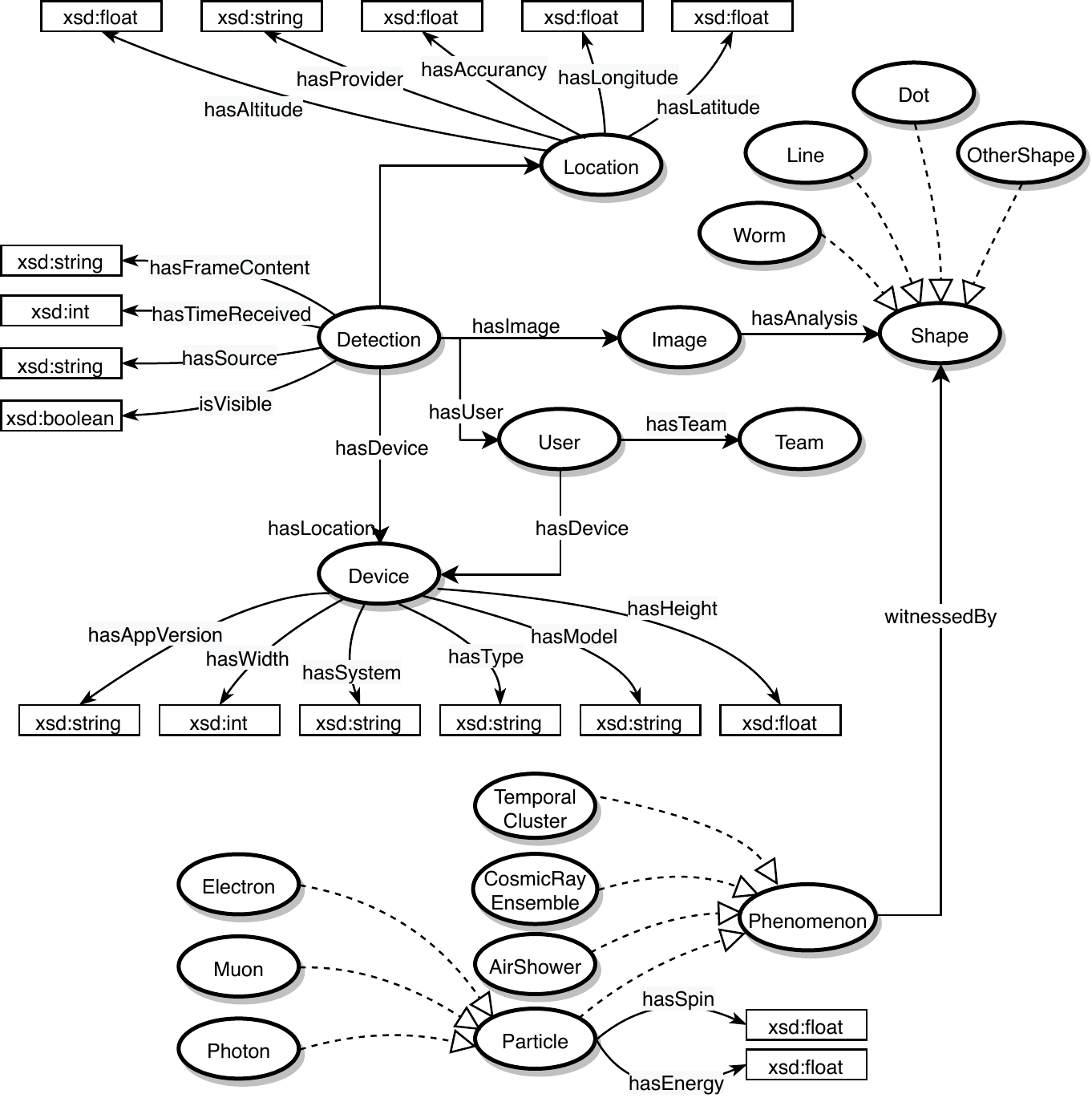}
\caption{\label{fig:ontology}Simplified schema of the CREDO Ontology. Classes appear as ovals, data types appear as boxes. Properties of classes appear as solid
arrows, while the subclass relation appears as dashed arrows with triangular heads. For legibility reasons the prefix \qn{credo}{} is omitted from classes and
property names.}
\end{figure}

The CREDO Ontology
\cite{credo_ontology} is the first step towards making CREDO data available through the LOD network.
The CREDO ontology is a lightweight OWL ontology. The current version
 features 20 classes, 14 object properties (that represent relations between
classes) and 40 data properties (that represent data attributes). A simplified schema of the ontology is depicted in Figure~\ref{fig:ontology}.

One part of the classes provides schema for the raw detection data that is
produced by the CREDO Detector app and other sources and is currently available
at the CREDO servers. Instances of the \qn{credo}{Detection} class capture data
associated with each detection. Some of these are directly expressed using data
properties, however the data about the user, their device, the image captured
and the location of the detection are associated as instances of the respective
class \qn{credo}{User}, \qn{credo}{Device}, \qn{credo}{Location}, and
\qn{credo}{Image}.

In addition, the ontology features classes that do not directly correspond to raw data recorded during detections, but rather they are intended to capture information derived from the analysis of these data. This includes the class \qn{credo}{Shape} and its four (mutually disjoint) subclasses \qn{credo}{Dot}, \qn{credo}{Line}, \qn{credo}{Worm}, and \qn{credo}{OtherShape}. These are used to classify the artefacts found in the detection images based on their geometric shape analysis. 

Finally, if scientific analysis of the detected artefacts, their time frames and shapes identify that they signify a manifestation of a physical phenomenon, such phenomena can be captured under the \qn{credo}{Phenomenon} class and its subclasses, for example \qn{credo}{Particle}, \qn{credo}{AirShower}, \qn{credo}{TemporalCluster}, and others.

While the CREDO data is currently available in a non-semantic format via the CREDO API as described above, the full utility of the CREDO ontology will be reached once the data is also available in the semantic format via a SPARQL endpoint. This is part of our ongoing efforts.

\section{Building the scale: public engagement as a scientific need}
\label{sec-citizens}

\vspace{\baselineskip}
\begin{justify}
As explained in more details in Ref. \cite{Homola2019-xo}, since a planetary scale and massive, geographically spread data acquisition with even small detectors such as smartphones are essential for the CREDO mission, the CREDO Collaboration considers an active engagement in the project of non-expert science enthusiasts as a methodological must: beneficial both for the project and for the contributors, and making science experts and non-professionals just one community, working hand-in-hand towards common goals. The unprecedented cosmic-ray data set to be collected and processed by CREDO will require a continuous overlook and vigilance of many people. Thus public engagement must be planned on many levels: from passive data taking with smartphones, through simple data monitoring, mining and analysis via the Internet, to more advanced activities. In practice, it means that CREDO will offer long-lasting educational and developmental paths for individuals who became amazed after their first, very close and active contact with science, and who will be ready for more. If the scale of the CREDO network is as large as planned, our format must bring benefits not only to the whole science community but also to the society. Universally useful educational and mind formation opportunities together with the availability of indirect and immediate participation in a top-science project for the so-far-excluded countries, regions and individuals, should contribute to the overall, sustainable civilizational development of society.
\end{justify}

\begin{figure}[H]
	\begin{Center}
		\includegraphics[width=0.9\textwidth]{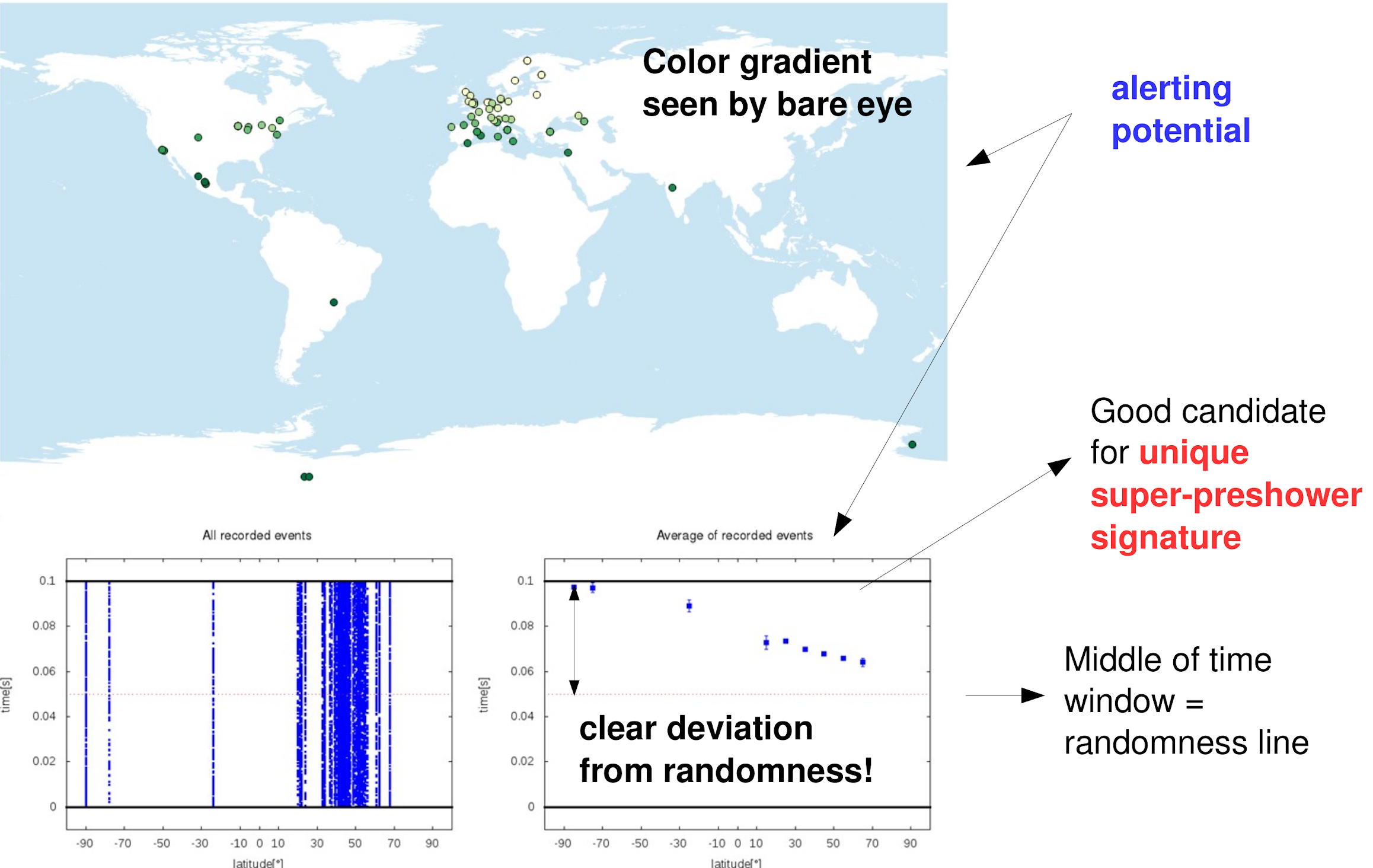}
	\end{Center}
	\caption{The concept of the CRE global pattern detection/alerting. See the text for details. \cite{Homola2019-xo}}
	\label{fig-duw}
\end{figure}

\begin{justify}
One of the basic (very simple) trigger concepts related to CRE and public engagement potential concerns global cosmic ray patterns, is illustrated in Fig. 10, using an example toy CRE simulation (arbitrarily dense and wide particle front arriving to the Earth from geographical North) and some random noise. The simulations are thrown on top of a global network of cosmic-ray detectors which, for simplicity, detect all the particles. One sees that a strong enough CRE signal can give a visible, global pattern built of the stations that see particles within some predefined time window. The pattern can be visualized by putting the locations of the active stations on the globe map and color-coding average arrival times of all particles hitting the detectors (top panel of Fig. \ref{fig-duw}). One can also show the arrival time scatter plot (bottom-left) or/and average arrival times for the stations located within a certain territory (bottom-right -- here within the 10 degree wide latitudinal belts). A random pattern is expected to give the average values located, within the uncertainties, in the middle of the considered time window, while the CRE-like signal should be visible as a departure from the ``randomness line''. So the pattern classification is based on a distinction between ``flat line in the middle of the window'' and ``deviations from the flat line in the middle''. This fundamental simplicity is then similar e.g. to the popular citizen science project designed to search for new planets, Planet Hunters \cite{Fischer2012-dl}, where the light curves of stars are exposed to a crowd-sourced classification and the users are asked to look for ``transits''  visible as periodic reduction of the received star light. This kind of pattern recognition is of course easily performed with simple algorithms or neural networks, nevertheless humans are still needed to classify the images ``on the edge'' or ``super strange'' -- seen for the first time and thus obviously not included in the training sets. In Planet Hunters the number of images suitable for human-based classification is large, and one expects even larger data load in the CREDO citizen science interfaces dedicated to global patterns: Private Particle Detective (still in the prototype phase) Dark Universe Welcome \cite{Dark_univ}, both powered by  Zooniverse \cite{zooniverse}, the platform which also hosts Planet Hunters. Thus in both cases the involvement of a large number of users is not just a funny outreach option, it is a must of fundamental importance: there might be only a few really important observations (the most Earth-like planets in the case of Planet Hunters or the clearly non-random super-preshowers in the case of Dark Universe Welcome) and we do not want to miss them. All hands (smartphones, other detectors) are needed on board.  
\end{justify}


\begin{justify}
Apart from the central public engagement effort there is and will be a number of 
peripheral and potentially beneficial activities initiated
by both professionals and non-expert participants of the CREDO program. The room for having and developing one's own initiatives and ideas that can be potentially available and useful to the whole CREDO community has already been proven to serve as an efficient public attractor. Example results of non-expert peripheral activities related to CREDO include e.g. Windows/Linux applications using cameras as radiation detectors \cite{CREDO-PC-Window,Credo-detector-for-linux}, or a set of Python and SQL tools for basic data visualization and analysis \cite{Windows-Tools,PyCharm-scripts} (see an example screenshot in Fig. \ref{fig-mpknap}).


\begin{figure}[H]
	\begin{Center}
		\includegraphics[width=0.9\textwidth]{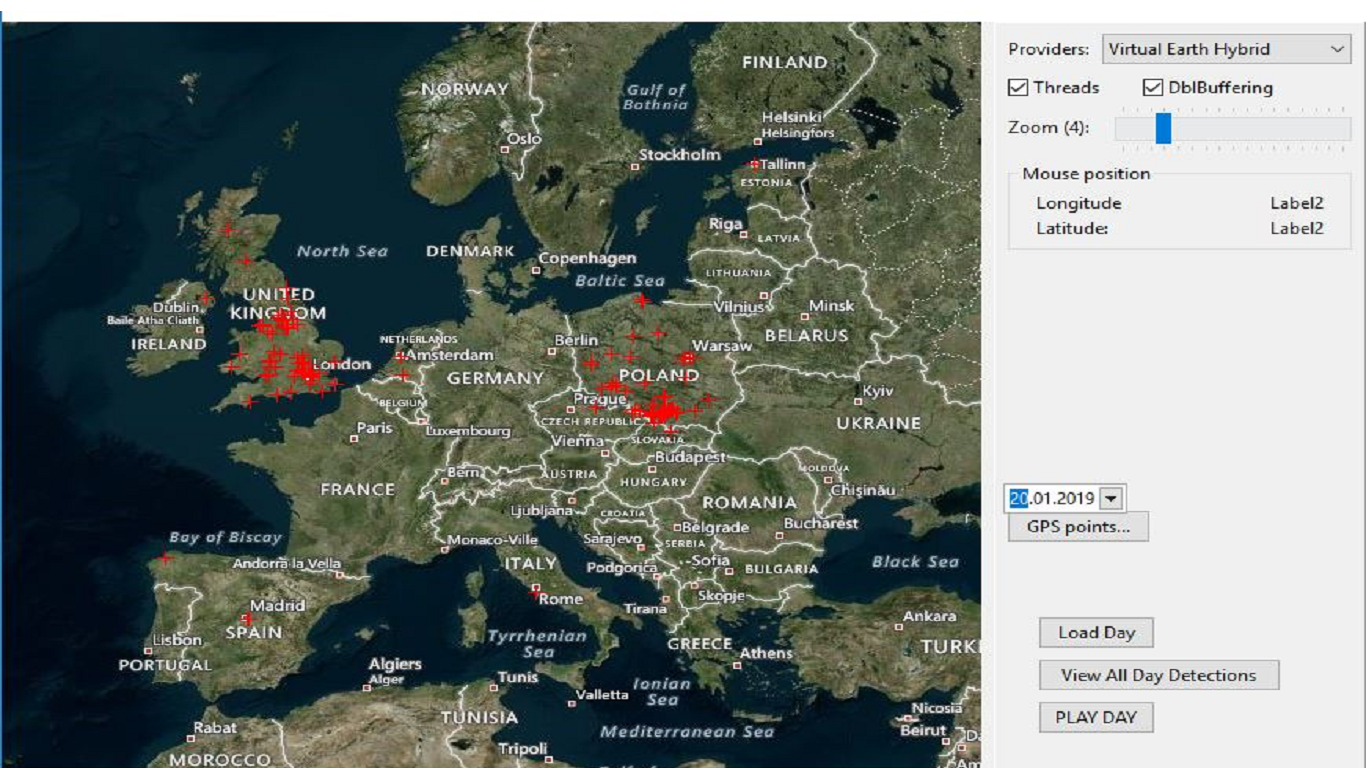}
	\end{Center}
	\caption{A basic visualization of the geographical locations of CREDO detections performed by a script developed by one of the non-expert CREDO participants.}
	\label{fig-mpknap}
\end{figure}


A very promising direction, both in terms of data acquisition and public engagement, is a detector application that can be run using a web browser independently of the operating system of the device. The project named ``CREDO-web-detector'' is now being developed by the students of Cracow University of Technology. It is an attempt to recreate the Android version of the CREDO Detector mobile application in a web browser environment.
Web browsers offer a good opportunity to reach new participants, as their implementations are present in all popular operating systems. If a web-browser-based detector application is used in future, CREDO can be universally opened to Microsoft Windows, Apple IOS or Linux users.
The trial version of CREDO-web-detector application is available in the public CREDO software 
repository \cite{credo-web-detector}.
\end{justify}

\begin{justify}
Another way of public engagement stimulation and improving the user retention rate that has already been implemented in CREDO is based on introducing gamification and edutainment components into scientific projects. For example, smartphone data acquisition and particle identification studies are noticeably supported by the prototype team competition in particle detection named Particle Hunter League \cite{hunters}. During the initial phase of operation advertised mostly among teachers in southern Poland the competition attracted around 2000 pupils from $\sim$100 teams (schools), and it is now being planned to internationalize the format beginning from the 17 CREDO member countries.
\end{justify}

\begin{justify}
Apart from a set of easy access solutions CREDO offers also higher level opportunities like virtual summer practices, supervision of bachelor, master and PhD projects, and also openness of the scientific teams dedicated to specific tasks reflecting the central CREDO agenda. More advanced CREDO participants and users oriented mostly on software development can also benefit from the fact that CREDO operates using open source codes under the MIT license \cite{mit}. This kind of openness enables healthy competition among the teams of software developers in a hackathon-like style, leading hopefully to a ``natural'' evolution of the software solutions within the CREDO program, although to make use of such non-central software activities a regular quality and functionality assessment will have to be implemented centrally, with redirecting the most promising developers and their work to the further stages, with larger scale funding. 
\end{justify}

The above highlights and examples well illustrate the societal ambition of CREDO which can directly be inferred from the deep and novel sense of the scientific research described above, where the active and massive participation of the public is rather a necessity than an option. Based on how the first CREDO users reacted to our concept, and how hungry they were to participate in a deeply important scientific project as meaningful members – we are ambitious to ignite a new era in science based on a deep public engagement in conducting a scientific research, by using smartphones in the quest for understanding the Universe to the deepest, fundamental level, hence bringing the worlds of science professionals and ``just'' enthusiasts closer to each other than ever.

\section{Outlook}
\label{sec-outlook}


\begin{figure}[H]
\begin{center}
\includegraphics[width=0.95\textwidth]{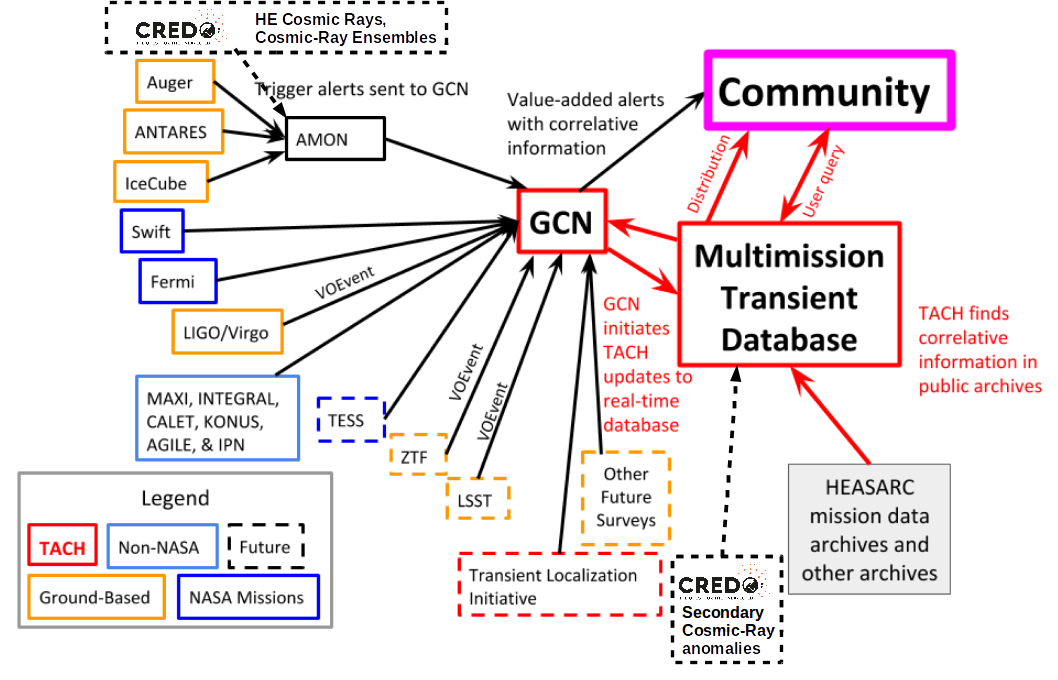}
\caption{The CREDO potential contributions to the Time Domain Astronomy Coordination Hub (TACH), a new NASA initiative. The CREDO logo has been positioned in two distinct places on top of the slide by Judith Racusin, NASA, from her invited talk \cite{multi-messenger} at the New Era of Multi-Messenger Astrophysics Conference, Groningen, March 2019.
}
\label{fig-tach}
\end{center}
\end{figure}

\begin{justify}
The current status of the Cosmic Ray Extremely Distributed Observatory described above as well as the dynamics of the progress being made by the CREDO Collaboration allows an optimistic outlook regarding the future perspective of further development and the resulting accomplishment of the ambitious scientific objectives. However, the apparently bright and scientifically exciting future of the CREDO mission still requires attention and care today. Apart from the already functional, fast growing own original data acquisition system which provides first scientific results the CREDO concept needs to attract more intensely the wide astroparticle physics community to make an optimum use of the available infrastructures, data reservoirs and expertise resources -- as must be required by any global approach to the observation and studies on cosmic ray ensembles. This need naturally does not imply any reorientation or renaming of the existing scientific collaborations and programs, it just defines a chance for synergy leading to new observations in the field of astroparticle physics within an unprecedentedly wide energy regime, including the highest energies known. One of the pillars of the awaited success is tightly connected with the principles of Open Science \cite{Woelfle2011-qo} focused on making scientific research (including publications, data, physical samples, and software) and its dissemination accessible to all levels of an inquiring society. One of the advantages of the Open Science research is a potential contained in the inter-collaborative effort dedicated to multi-messenger and multi-mission projects, reflected e.g. in the recent NASA initiative named Time Domain Astronomy Coordination Hub (TACH) \cite{Smale2020-ih}, see Fig. \ref{fig-tach}. CREDO directly addresses the TACH perspective, although the possible inter-cooperative activities might have different faces. One of the example directions that have been already explored is based on the expectations the gamma rays collected by CTA might be components of Cosmic-Ray Ensembles that CREDO has argued in the aforementioned Ref. \cite{Almeida_Cheminant2020-ut}, and there might me more direct links to the other ESFRI \cite{ESFRI} infrastructures. For example, the unwanted muon background received by KM3NeT \cite{Kouchner2017-yl} (or any other underwater neutrino observatory!) might serve as a signal for the CRE-oriented strategies, and E-ELT \cite{Gilmozzi2007-fi} (and other telescopes as well!) can contribute to CREDO by sharing their unwanted cosmic-ray ``artifacts'' recorded with the CCD cameras in the ``dark frame'' mode or during regular operation. And these are only selected examples given to illustrate the ``connectivity''  of the CRE research. We indicate places where the planned CREDO multimessenger and multimission contributions can be made on top of the NASA perspective in Figure \ref{fig-tach}: 1) high energy cosmic rays and cosmic-ray ensemble alerts provided by CREDO can naturally contribute to TACH via the Astrophysical Multimessenger Observatory Network (AMON) \cite{amon} dedicated to high energy astroparticle physics phenomena, and 2) the alerts on transient anomalies in secondary cosmic-ray radiation detection rate can be sent directly to the Multimission Transient Database. CREDO will of course benefit also from receiving alerts from TACH which will expand and enrich the scope of scientific research opportunities (enabling e.g. also studies on the sub-threshold level) available to the CREDO participants and users. The planned connection to AMON, and further to the TACH network, will give an opportunity to further enhance the credibility of the CREDO data, increase its scientific potential, and stimulate the CREDO infrastructure development adequately to the expected scientific and societal interest. 
\end{justify}

\begin{justify}
It is worthwhile to stress that the multi-mission CREDO potential might connect the astronomy and astroparticle physics communities with other fields and create inter- and transdisciplinary opportunities that might turn out to be of vital everyday life importance for a large fraction of humanity. We list here a potential for studying correlations of transient anomalies in detection rates of low energy secondary cosmic rays with seismic effects. The search for such correlations will be carried out based on multi-messenger signals received by global networks of detectors, including cosmic-ray sensitive devices operated by CREDO. Mass movements inside the Earth that lead to earthquakes simultaneously cause temporary changes in the gravitational and geomagnetic field. These changes are propagating at the speed of light and can therefore perhaps be observed on the surface earlier than earthquakes, for example by registering changes in the frequency of detection of cosmic radiation, which is very sensitive to geomagnetic conditions. Confirming of such correlations would give hope for a warning system, and an effort in this direction has a motivation described in the literature \cite{Morozova2000-wu,Foppiano2008-xo,Romanova2015-ci,Pisa2010-tm,2018JGRA_He}, however no infrastructure suitable for global studies has been available.  
\end{justify}

\begin{justify}
Another interesting interdisciplinary line of work with CREDO is Space Weather (and Space Climate over long periods of time). Besides a lot of possibilities for specialists, CREDO can be used as an excellent introduction to space weather for the general public since cosmic rays are a fundamental part of the space environment. Until now, the classic experiment that served as an introduction to space weather for the general public was the counting of sunspots with a modest telescope \cite{2019NatAs_MJV}. However, the generalization of mobiles makes CREDO an ideal instrument to engage the general public interested in Space Weather.
\end{justify}

\begin{justify}
Another interdisciplinary opportunity is global monitoring of the locations and arrival times of extensive air showers. Average doses from cosmic ray radiation are known to be significantly lower than those due to local and natural radiation, but it is yet unknown what impact on biological systems comes from irradiation by very energetic EAS, especially their central regions. 
Particle mass composition, density and energy distributions there contribute to a single radiation dose which is hard or infeasible to mimic in laboratories, and thus not available for direct reflecting/projecting in basic laboratory research.
The CREDO infrastructure will enable this sort of pioneer investigation by connecting geographic coordinates and arrival times of energetic EAS cores with the locations of humans engaged in the project, providing the data for long-term studies on possible correlations between cosmic radiation and our health. Such studies will potentially lead to explaining the causes of some diseases of unknown aetiologies, with hope for new therapies and/or preventive examination strategies. This interdisciplinary research will go beyond the existing research frameworks both in the fields of the physics of cosmic radiation itself and concerning the knowledge about the possible biological response of the human body to this radiation. Our studies will be conducted in accordance with the global trends in understanding the impact of low doses of radiation on living organisms in terms of the possibility of the occurrence of both positive phenomena (e.g. increasing the immunity of organisms) and negative ones (e.g. some types of cancer) \cite{Kreuzer2018-xs}. These studies may shed light on interesting facts that still cannot be explained as the strong inverse correlation found in Italy between high solar activity and incidence of schizophrenia and bipolar disorder \cite{2011AdSpR_VB}.
\end{justify}\par

\section{Summary and Conclusions}
\label{sec-sum}
\begin{justify}
With CREDO we are going to explore a new observation channel of the Universe, the CRE channel. The width and complexity of the scope of the studies point to a long-term research perspective, and to the need of engaging wide communities of both professional scientists and non--expert science enthusiasts. The outcome of this program will tell whether the CRE channel broadcasts with New Physics or ``just'' new upper limits to astrophysical and cosmological scenarios based on null CRE results. If CRE are observed, they could point us back to the interactions at energies close to the GUT scale and lead us to a breakthrough in physics. If CRE are not observed, we will set the new upper limits constraining selected theories, e.g. the SHDM models, which will point to a new way to narrow the search for a breakthrough in science. You are welcome to join us!
\end{justify}



\authorcontributions{
conceptualization, P.H., T.W.; 
methodology,  D.Be., N.D., P.G., K.G., M.H., P.H., M.Kar., R.K., S.K., M.M., J.W.M., A.N., M.N., C.O., K.O., M.Pa., K.Sme., O.S., T.W.; 
software, \L{}.Bi., M.B., D.Bu., A.\'C., N.D., A.R.D., P.G., M.Kn., S.K., M.M., J.M\k{e}., M.N., K.O., M.Pa., M.Pi., B.P., S.S., O.S.; 
validation, M.B., P.H., J.M\k{e}., B.P., S.S., O.S.; 
formal analysis, J.P., W.S., T.W., K.W.W.; 
investigation, M.B., K.A.C., N.D., P.H., J.M\k{e}., B.P., J.P., W.S., S.S., O.S., T.W., K.W.W.; 
resources, P.H., M.Kn., M.M., M.N., B.O., K.O., M.Pa., M.R., J.Su.; 
data curation, N.D., M.H., Z.H., O.S., S.S.; 
data science, \L{}.Bi., M.Pi., K.Rz.; 
theoretical studies, \L{}.Br., D.E.A.C., N.D., D.G., J.J., M.V.M., Y.J.N, J.Z.-S., A.T.; 
writing--original draft preparation, D.Be., \L{}.Br., D.E.A.C., K.A.C., N.D., P.G., K.G., D.G., A.C.G., J.J., S.K., M.H., P.H., R.K., M.Kn., M.V.M., J.W.M., A.N., C.O., K.O., M.Pa., J.P., K.R., J.Z.-S., K.Sme., W.S., S.S., O.S., A.T., T.W., K.W.W.; 
writing--review and editing, D.Be., G.B., \L{}.Br., N.B., K.A.C., D.E.A.C., P.D., A.R.D., K.G., A.C.G., M.H., P.H., M.Kas., R.K., P.K., B.\L{}., A.M., M.V.M., J.Mi., V.N., Y.J.N, B.O., G.O., M.Pa., M.R., K.R., J.Z.-S., K.Sme., K.Smo., J.St., S.S., O.S., M.S., A.T., K.T., J.M.V., T.W.; 
visualization, M.Kn., S.S., Z.H.; 
supervision,  \L{}.Bi., A.R.D., P.H., D.G., R.K., J.W.M., M.R., K.R., S.S., J.Su., T.W., K.W.W.; 
project administration, P.H., R.K., M.M., S.S.; 
funding acquisition,  M.H., R.K., A.T.
}

\funding{This research was partly funded by the International Visegrad grant No. 21920298. N. Budnev acknowledges partial support from The Russian Federation Ministry of Education and Science (Tunka shared core facilities -- unique identifier RFMEFI59317X0005), J. Zamora-Saa acknowledges support by FONDECYT (Chile), grant No. 3180032, M.V. Medvedev acknowledges partial support via the NSF grant No. PHY-2010109, the DOE EPSCOR grant No. DE-SC0019474 and the DOE grant No. DE-SC0016368, and Martin Homola is supported by Slovak VEGA grant No. 1/0778/18.  J.M. Vaquero acknowledges support from Junta de Extremadura-Fondo Social Europeo (GR18097).
}


\acknowledgments{We acknowledge the leading role in the CREDO Collaboration and the commitments to this research made by the Institute of Nuclear Physics Polish Academy of Science. This research has been supported in part by PLGrid Infrastructure and we warmly thank the staff at ACC Cyfronet AGH-UST for their always helpful supercomputing support. We thank Steven Carlip, Roger Clay, Bohdan Hnatyk, Qingdi Wang, and Henryk Wilczy\'nski for valuable remarks and discussions.}

\conflictsofinterest{The authors declare no conflict of interest.} 

\abbreviations{The following abbreviations are frequently used in this manuscript:\\

\noindent 
\begin{tabular}{@{}ll}
CR & Cosmic Ray(s) \\
CRE & Cosmic Ray Ensemble(s) \\
CREDO & Cosmic Ray Extremely Distributed Observatory\\
EAS & Extensive Air Shower(s) \\
UHECR & Ultra-High Energy Cosmic Rays
\end{tabular}}


\reftitle{References}


\externalbibliography{yes}
\bibliography{biblio_sum.bib}



\end{document}